\documentclass[useAMS,usenatbib]{mn2e}
\usepackage{subfigure}
\usepackage{ulem}
\usepackage{color}
\usepackage{graphicx}
\usepackage{times}
\usepackage{amssymb}
\usepackage{amsmath}
\usepackage{lscape}
\usepackage{url}
\usepackage{multirow}
\newcommand{\km}{\rm\thinspace km}

\newcommand{\cm}{\rm\thinspace cm}

\newcommand{\s}{\rm\thinspace s}

\newcommand{\Hz}{\rm\thinspace Hz}
\newcommand{\Msun}{\hbox{$\rm\thinspace M_{\odot}$}}

\newcommand{\erg}{\rm\thinspace erg}

\newcommand{\ergpcmsqpspA}{\hbox{$\erg\cm^{-2}\s^{-1}$\AA$^{-1}\,$}}

\newcommand{\ergpspHz}{\hbox{$\erg\s^{-1}\Hz^{-1}\,$}}
\newcommand{\kmps}{\hbox{$\km\s^{-1}\,$}}

\newcommand{\psqcm}{\hbox{$\cm^{-2}\,$}}

\newcommand{\ps}{\hbox{$\s^{-1}\,$}}

\newif\ifAMStwofonts
\AMStwofontstrue
\def\Lya{Ly$\alpha$~}

\def\LCDM{$\Lambda$CDM~}
\def\HI{\hbox{H$\,\rm \scriptstyle I\ $}}

\def\HeII{\hbox{He$\,\rm \scriptstyle II\ $}}

\def\MgII{\hbox{Mg$\,\rm \scriptstyle II\ $}}
\def\NV{\hbox{N$\,\rm \scriptstyle V\ $}}

\begin{document}

\title[Proximity effect measurements of the UVB at high redshift]
      {Measurements of the UV background at $\bmath{4.6 < z < 6.4}$
      using the quasar proximity effect
      \thanks{A part of the observations were made at the W.M.~Keck
      Observatory which is operated as a scientific partnership
      between the California Institute of Technology and the
      University of California; it was made possible by the generous
      support of the W.M.~Keck Foundation. This paper also includes
      data gathered with the 6.5 meter Magellan Telescopes located
      at Las Campanas Observatory, Chile.}
      }
\author[A.P.~Calverley et al.] 
{\parbox[]{6.in}
   { Alexander P.~Calverley$^{1}$
   \thanks{E-mail: acalver@ast.cam.ac.uk},
   George D.~Becker$^{1}$,
   Martin G.~Haehnelt$^{1}$ and
   James S.~Bolton$^{2}$\\ } \\
   \footnotesize
   $^{1}$Kavli Institute for Cosmology and Institute of Astronomy, Madingley Road, Cambridge, CB3 0HA\\
   $^{2}$School of Physics, University of Melbourne, Parkville, VIC 3010, Australia}
   
\date{Accepted 2010 November 23. Received 2010 November 08; in original form 2010 September 13}

\maketitle

\begin{abstract}

We present measurements of the ionising ultraviolet
background (UVB) at $z \sim 5-6$ using the quasar
proximity effect. The fifteen quasars in our sample
cover the range $4.6 < z_{\rm q} < 6.4$, enabling the
first proximity effect measurements of the UVB at
$z > 5$. The metagalactic hydrogen ionisation rate,
$\Gamma_{\rm bkg}$, was determined by modelling the
combined ionisation field from the quasar and the UVB
in the proximity zone on a pixel-by-pixel basis. The
optical depths in the spectra were corrected for the
expected effect of the quasar until the mean flux in
the proximity region equalled that in the average \Lya
forest, and from this we make a measurement of
$\Gamma_{\rm bkg}$. A number of systematic effects
were tested using synthetic spectra. Noise in the
flux was found to be the largest source of bias at
$z \sim 5$, while uncertainties in the mean transmitted
\Lya flux are responsible for the largest bias at
$z \sim 6$.  The impacts of large-scale overdensities
and Lyman limit systems on $\Gamma_{\rm bkg}$ were also
investigated, but found to be small at $z > 5$. We find
a decline in $\Gamma_{\rm bkg}$ with redshift, from
$\log(\Gamma_{\rm bkg}) = -12.15 \pm 0.16$ at $z \sim 5$
to $\log(\Gamma_{\rm bkg}) = -12.84 \pm 0.18$ at $z \sim 6$
($1 \sigma$ errors). Compared to UVB measurements at lower
redshifts, our measurements suggest a drop of a factor of
five in the \HI photoionisation rate between $z \sim 4$
and $z \sim 6$. The decline of $\Gamma_{\rm bkg}$ appears
to be gradual, and we find no evidence for a sudden change
in the UVB at any redshift that would indicate a rapid
change in the attenuation length of ionising photons.
Combined with recent measurements of the evolution of the
mean free path of ionising photons, our results imply 
decline in the emissivity of ionising photons by roughly a
factor of two from $z \sim 5$ to 6, albeit with significant
uncertainty due to the measurement errors in both
$\Gamma_{\rm bkg}$ and the mean free path.

\end{abstract}


\section{Introduction}
\label{sect:intro}
The metagalactic ultraviolet background (UVB) is the net
radiation field responsible for keeping the Universe
ionised from the end of reionization to the present day.
The relative contributions from galaxies and quasars, as
well as filtering by the intergalactic medium (IGM)
itself, determine the intensity and spectrum of the UVB
\citep[e.g.][]{Bechtold1987a,Haardt1996a,Fardal1998a, Haardt2001a}.
Thus by measuring the UVB one can hope to place constraints
on the evolution of the source population with redshift.

Of particular interest is the evolution of the UVB at
$z \sim 6$. The appearance of Gunn-Peterson (GP) troughs
in the spectra of the highest-redshift known quasars has
been interpreted as evidence for a sharp downturn in the
UVB at $z > 6$ signalling the end of reionization
\citep[e.g.,][]{Fan2006a}. However, the diminishing
transmitted flux is also consistent with a more slowly-evolving
UVB and IGM density field \citep{Bolton2007a,Becker2007a}.

The bulk of the ionising photons at $z>6$ that make up
the UVB are believed to come from low-luminosity galaxies  
\citep{Richard2006a,Stark2007b,Richard2008a,Srbinovsky2010a,Oesch2010a}.
Direct searches for these sources at $z \gtrsim 6-10$ have
taken advantage of recent very deep optical and near-infrared
imaging from both the ground and space
\citep{Bunker2004a,Bouwens2006a,Yoshida2006a,Bouwens2008a,Ouchi2009a}.
The majority of these faint galaxies, however, still remain
below current detection thresholds \citep[e.g.][]{Bouwens2010a}.

The quasar proximity effect has been a classic tool for
directly measuring the intensity of the UVB at high
redshifts. Since quasars are highly luminous, their
output of ionising photons will dominate over that of
the UVB out to large (up to several proper Mpc) distances.
This produces a region of enhanced transmission near the
redshift of the quasar, first noted by \citet{Carswell1982a},
known as the `proximity region.' The size of this region
depends both on the quasar luminosity and the intensity
of the UVB. For a known quasar luminosity, therefore,
the UVB can be estimated by measuring the extent of the
proximity zone. Classically, the proximity effect has been
measured by comparing column densities of the \HI \Lya
absorption lines in the forest with those close to the quasar
\citep[][hereafter BDO]{Murdoch1986a, Tytler1987a, Carswell1987a, Scott2000a, Bajtlik1988a},
although a variety of flux statistics have also been used
\citep{Liske2001a,DallAglio2008a,DallAglio2010a}.
Altogether, proximity effect studies have delivered
measurements of the UVB from $z \sim 0.5$ \citep{Kulkarni1993a}
to $z \sim 4.5$ \citep{DallAglio2009b}.

More recently, an alternative method of estimating the UVB
has been developed which uses the mean flux in the \Lya
forest in combination with numerical simulations. The UVB
in the simulation is adjusted until the mean flux in artificial
\Lya forest spectra is equal to that in the real data
\citep[see e.g.,][]{Rauch1997a,Songaila1999a,Tytler2004a,Bolton2005a,Kirkman2005a,Jena2005a}.
At $z > 4.5$, the UVB has so far only been determined using
this method
\citep{McDonald2001a,Meiksin2004a,Bolton2007a,Wyithe2010a}.
At $z \sim 5-6$, however, converting the mean flux into an
ionisation rate depends sensitively on modelling the gas density
distribution at very low densities \citep[e.g.][]{Miralda2000a}.
Not only is this a challenge numerically \citep{Bolton2009a},
but the optical depth distribution will depend on the properties
of the simulation, including the gas temperature.

At $z < 4.5$, estimates of the UVB from the proximity effect
and flux decrement methods can be directly compared. These
are generally discrepant at $2<z<4$, with the proximity effect
estimates systematically higher for most studies. Recent
proximity effect papers suggest that there is a competition
between the effect of the enhanced intensity of the UVB and
the overdensity of matter close to the quasar. Classically,
the proximity effect assumes that the density distribution
close to the quasar is the same as that in the general IGM.
Not accounting for overdensities may therefore cause the UVB
to be overestimated by up to a factor of 3 \citep{Loeb1995a}.
By using an independent measure of the UVB from flux decrements,
\citet{Rollinde2005a} found tentative evidence that quasars
at $z=2-3$ may reside in haloes as massive as $\sim 10^{14}\Msun$,
similar to the most massive halo in the Millennium simulation.
\citet{Guimaraes2007a} have claimed to detect a similar effect
at $z \sim 4$. Other recent papers, however, have claimed to
have overcome the environmental bias of an enhanced average
density in the proximity zone. \citet{DallAglio2008a} report
that with their measurement method only 10 per cent of their
sample at $2 \lesssim z \lesssim 4.5$ showed significant
excess absorption attributable to an overdense environment,
and that for the majority of the quasars in their sample
their proximity effect measurement of the UVB did not appear
to be affected by an overdense quasar environment (at least on
scales $>3$ Mpc).

In this paper, rather than identify individual lines, which
becomes increasingly difficult at $z > 4$, we further develop
a variant on the `flux-transmission' method \citep{Liske2001a}
to measure the UVB intensity at $z > 4.5$. At lower redshifts
this has been used to compare the mean flux averaged over
extended sections of the spectrum near the quasar redshift to
the mean flux in the forest. Rather than compute the mean
flux in sections, we consider individual pixel optical depths
across the proximity region. The optical depths are modified
to remove the presumed effect of the quasar, until the
proximity region has the same mean flux as the forest at that
redshift. The characteristic scale length of the quasar model
is then combined with the quasar luminosity to estimate the UVB. 
This simple approach avoids a direct dependence on simulations
as it does not require the optical depth distribution of the
forest to be known {\it{a priori}}. We use, however, simulations
extensively to estimate the bias and uncertainties of our method.

The remainder of this paper is organised as follows:
In section 2 we describe the observational data as well as the
hydrodynamical simulations used. In section 3 we detail the
proximity effect analysis, and the sources of systematic bias
such as the effect of the quasar environment. The results are
presented and their implications discussed in Section 4.
Finally, we present our conclusions in Section 5. Throughout
this paper we assume a flat Universe and cosmological parameters
taken from the mean of the WMAP 5 year data set \citep{Komatsu2009a},
with Hubble constant H$_0$ = $72~ \rm kms^{-1}~Mpc^{-1}$ and
density parameters ($\Omega_m$, $\Omega_\Lambda$) = (0.26, 0.74).

\begin{table*}
\centering
  \caption{The list of quasars included in this paper.
  Columns give the quasar name and redshift, details of the
  observations, and the average signal-to-noise per pixel in
  the $40h^{-1}$ comoving Mpc closest to the quasar redshift,
  after masking out skyline residuals (see Section~\ref{sect:reliable}).
  }

  \begin{tabular}{c|c|c|c|c|c|c}
    \hline
    \hline
    Name						&$z_{\rm q}$ & Inst. & Dates               &$t_{\rm exp}$ (hrs)& Ref. & $S/N$  \\
    \hline
    SDSS J1148+5251 & 6.42       & HIRES & Jan 2005 - Feb 2005 & 14.2$^{\rm a}$    & 1	  & 16   \\
    SDSS J1030+0524 & 6.31       & HIRES & Feb 2005            & 10.0              & 1	  & 12   \\
    SDSS J1623+3112 & 6.25       & HIRES & Jun 2005            & 12.5              & 1	  & 11   \\
    SDSS J0818+1722 & 6.02       & HIRES & Feb 2006            & 8.3               & 2	  & 12   \\
    SDSS J1306+0356 & 6.02       & MIKE  & Feb 2007            & 6.7               & 4	  & 14   \\
    SDSS J0002+2550 & 5.82       & HIRES & Jan 2005 - Jul 2008 & 14.2              & 1,4	& 21   \\
    SDSS J0836+0054 & 5.81       & HIRES & Jan 2005            & 12.5$^{\rm a}$    & 1	  & 19   \\
    SDSS J0231-0728 & 5.41       & HIRES & Jan 2005 - Feb 2005 & 10.0              & 1	  & 14   \\
    SDSS J1659+2709 & 5.33       & HIRES & Sep 2007 - Jul 2008 & 11.7              & 3	  & 32   \\
    SDSS J0915+4924 & 5.20       & HIRES & Feb 2005            & 10.0              & 1	  & 23   \\
    SDSS J1204-0021 & 5.09       & HIRES & Jan 2005 - Feb 2005 & 6.7               & 1	  & 17   \\
    SDSS J0011+1440 & 4.97       & HIRES & Sep 2007            & 6.7               & 3	  & 47   \\
    SDSS J2225-0014 & 4.89       & MIKE  & Oct 2007            & 5.0               & 4	  & 23   \\
    SDSS J1616+0501 & 4.88       & MIKE  & Mar 2008            & 3.3               & 4	  & 21   \\
    SDSS J2147-0838 & 4.59       & MIKE  & Oct 2007            & 8.3               & 3	  & 51   \\
   \hline
   \multicolumn{7}{|l|}{$^{\rm a}$ The present reductions include only data taken with the upgraded detector.} \\
   \multicolumn{7}{|l|}{References:} \\
   \multicolumn{7}{|l|}{1 - \citet{Becker2006a}; 2 - \citet{Becker2007a}; 3 - \citet{Becker2010a}; 4 - This paper} \\
\end{tabular}
\label{tab:targets}
\end{table*}

\section{Data and models}

\subsection{Observed spectra}
\label{sect:obs_spec}
The quasar spectra used in this paper were taken
with either the Keck or Magellan telescopes. The
majority of the quasars at $z > 5$ were observed
with the High Resolution Echelle Spectrometer
\citep[HIRES;][]{Vogt1994a} on the 10~m Keck I
telescope, and reduced using a custom set of IDL
routines and optimal sky subtraction techniques
as detailed in \citet{Becker2006a,Becker2007a}.
All of these observations were made with the 0.86"
slit with $R=40\,000$, and so the velocity
resolution is $6.7\kmps$.

The majority of the quasars at $z < 5$ were observed
using the 6.5~m Magellan-II Clay and the Magellan
Inamori Kyocera Echelle (MIKE) spectrograph
\citep{Bernstein2003a} and reduced with a similar
custom pipeline. The velocity resolution is roughly
half that of the HIRES spectra at $13.6\kmps$.
A list of the targets is shown in Table~\ref{tab:targets}.

Quasar redshifts were taken either from the CO and
\MgII redshifts presented in \citet{Carilli2010a},
or from the spectra themselves by identifying the
redshift at which the \Lya forest appears to start.
Errors on the redshifts measured from the apparent
start of the \Lya forest were estimated by
comparing to more precise redshifts from \MgII and
CO where available, or those in the SDSS. All the
objects have photometry in the SDSS, and those at
$z < 5.5$ also have flux-calibrated spectra in the
SDSS archive. Continuum magnitudes for those at
$z>5.5$ were taken from the discovery papers
\citep{Fan2001a, Fan2003a, Fan2004a, Fan2006a},
whilst fluxes for those at $z < 5.5$ were measured
from the SDSS spectra. In both cases the continuum
flux was measured at a rest wavelength of 1280~\AA.
The error in the flux measured from the SDSS
spectra is conservatively taken to be
$0.5 \times 10^{-17}\ergpcmsqpspA$.

All the spectra were normalised following the method
described in \citet{Bolton2010a}. The spectrum is
first divided through by a power-law
$F_{\nu} \propto \nu^{-0.5}$, normalised at
$1280\,(1+z)$~\AA, and the \Lya emission line is then
fitted with a slowly varying spline. It is difficult to
fit the continuum over the forest at these redshifts
due to the low flux levels, however in the proximity
region the transmitted flux maxima will be nearer to the
continuum, and so is estimated to be within $\sim 20$
per cent of the correct value over the region of interest.

\subsection{Simulated spectra}
\label{sect:sim_spec}

\begin{table}
\centering
  \caption{Mass resolution and box size (comoving) of the
    hydrodynamical simulations used in this work. Model C
    was primarily used to simulate quasar proximity zones.}

  \begin{tabular}{c|c|c|c}
    \hline
    \hline
    Model    & $L$     & Total particle         &  $M_{\rm gas}$  \\
             & [Mpc/h] &  number  							&  $[\rm M_{\odot}/h]$ \\
  \hline

  A     & 20     & $2 \times 100^{3}$   & $1.03 \times 10^{8}$ \\
  B     & 20     & $2 \times 200^{3}$   & $1.29 \times 10^{7}$ \\
  C     & 20     & $2 \times 400^{3}$   & $1.61 \times 10^{6}$ \\
  D     & 40     & $2 \times 200^{3}$   & $1.03 \times 10^{8}$ \\
  E     & 40     & $2 \times 400^{3}$   & $1.29 \times 10^{7}$ \\
  F     & 80     & $2 \times 400^{3}$   & $1.03 \times 10^{8}$ \\
   \hline
\end{tabular}
\label{tab:sims}
\end{table}

Simulations of the IGM at high redshift were used to
test the method and to explore potential sources of
systematic error. The simulations are the same as
those in \citet{Bolton2009a} and were performed using
a customised version of the parallel Tree-SPH code
{\tt{GADGET}}-3, an updated version of the publicly
available code {\tt{GADGET}}-2 \citep{Springel2005a}.
The simulations assume the cosmological parameters
$(h,\Omega_m,\Omega_\Lambda,\Omega_b h^2,\sigma_8) = (0.72,0.26,0.74,0.024,0.85)$,
have both dark matter and gas components, and were
started at $z=99$ with initial conditions generated with
the transfer function of \citet{Eisenstein1999a}. Each
simulation uses the UVB model of \citet{Haardt2001a} with
contributions from both galaxies and quasars, and is
switched on at $z=9$ and applied in the optically thin
limit. Our fiducial run is a $20h^{-1}$ (comoving) Mpc
box with $2 \times 400^3$ particles (Model C).
Simulations with other box sizes and particle numbers
were used to test the dependence of the environmental
bias due to an enhanced average density on the mass of
the host halo, and are summarised in Table~\ref{tab:sims}.

Simulated \Lya spectra were constructed from line-of-sight
density, peculiar velocity, neutral \HI fraction and
temperature fields. For the main analysis this was done for
1024 random sightlines drawn parallel to the box boundaries
\citep[e.g.][]{Theuns1998a}, with outputs at $z = (2,3,4,5,6)$.
Each sightline is 1024 pixels long. Proximity zones were
introduced into the spectra by modifying the neutral \HI
fraction (in real space) with the ionising intensity
falling off as $1/r^2$, before convolving with the
other fields to derive the optical depth (in velocity space).
The spectra were convolved with a Gaussian with FWHM equal
to the velocity resolution of the instrument that was
being modelled ($6.7\kmps$ for HIRES and $13.6\kmps$ for MIKE),
before being resampled at the instrument pixel resolution
($2.1\kmps$ and $5.0\kmps$ respectively). Gaussian distributed
noise, as well as other imperfections, could then be added.
Additionally, sightlines were drawn through the most
massive haloes for the analysis described in
Section~\ref{sect:halo-mass}. The haloes were identified
using a friends-of-friends algorithm with a linking length
of $0.2$. For the analysis in the rest of this paper
sightlines of $40h^{-1}$ comoving Mpc were used. For Models
A-C several random sightlines were combined for this.

\begin{figure*}
  \centering
    \includegraphics[width=18cm]{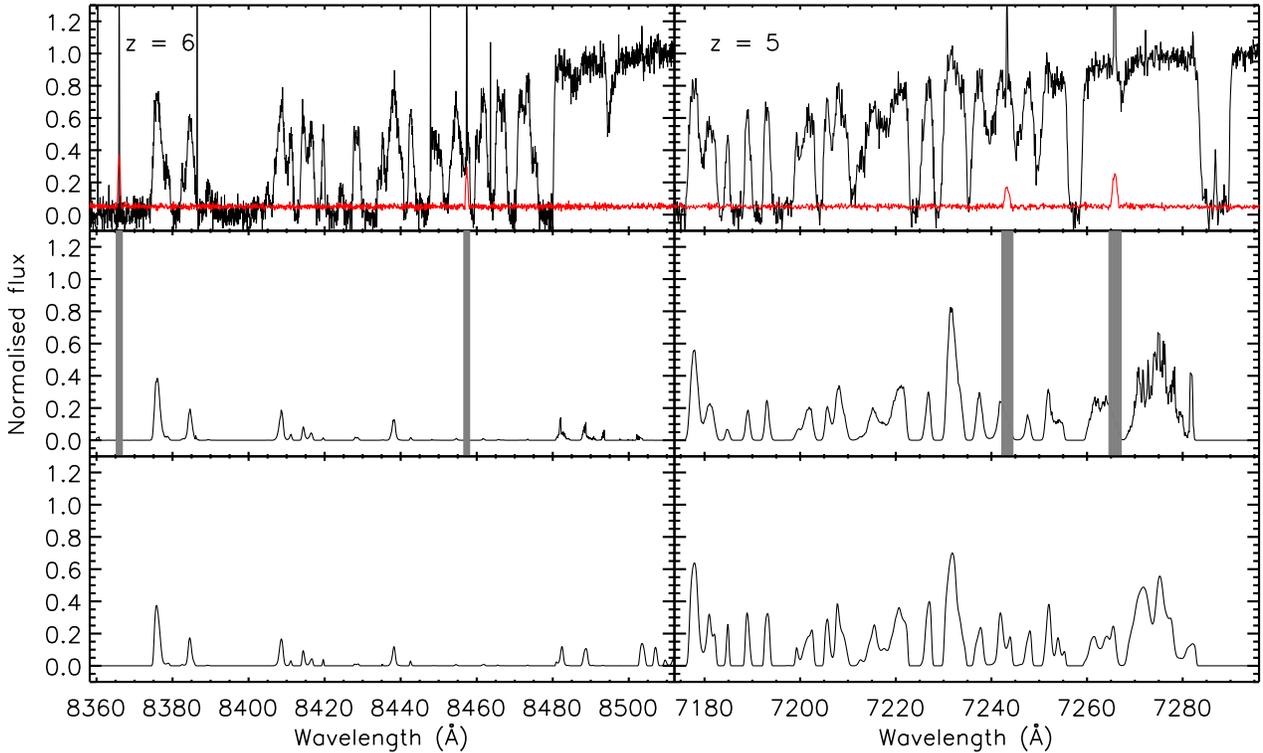}
  \caption{
  {\it{Top panels:}} Simulated sightlines at $z=6$ (left) and
  $z=5$ (right). Each sightline is $40h^{-1}$ comoving Mpc long,
  with $R_{\rm eq}$ = 10 proper Mpc and $S/N=20$. The resolution
  for the $z=6$ spectrum is equal to that of HIRES, whilst the
  spectrum at $z=5$ has resolution equal to that of MIKE. The
  red line is the simulated error array. Artifacts such as bad
  pixels and skyline residuals have also been introduced. \Lya
  at the redshift of the quasar is on the far right of each panel.
  {\it{Middle panels:}} The spectra smoothed according to their
  noise properties (see Section~\ref{sect:reliable}), and with
  optical depths altered such that $\Delta F =0$ (see
  Equation~\ref{delta_f-defn}). The grey regions mark parts of
  the spectra that were masked out automatically (primarily
  skyline residuals).
  {\it{Bottom panels:}} The original \Lya forest for these
  sightlines. The similarity between these spectra and those
  in the middle panels provides a good check that the method
  is working correctly.  
  }
  \label{fig:before_after}
\end{figure*}

\begin{figure*}
  \centering
    \includegraphics[width=18cm]{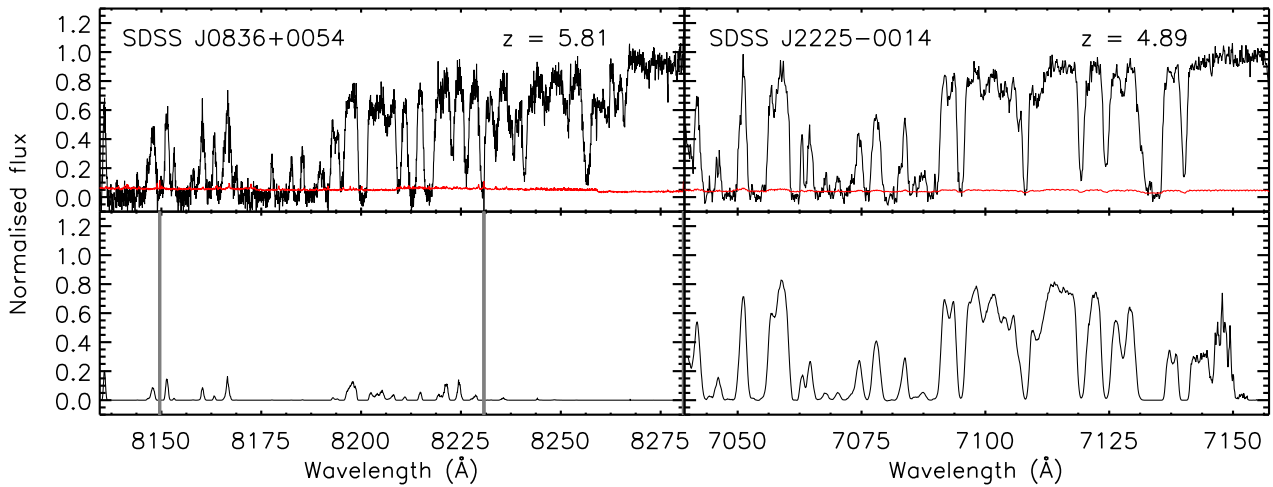}
  \caption{
  Same as Fig.~\ref{fig:before_after}, but for a small sample of
  the observed spectra.
  {\it{Top panels:}} The normalised spectrum of SDSS J0836+0054 at
  $z=5.810$ taken with the HIRES instrument (left), and the
  normalised spectrum of SDSS J2225-0014 at $z=4.886$ taken with
  MIKE (right). The length of each spectrum corresponds to
  $40h^{-1}$ comoving Mpc, and the red line is the observed error
  array. \Lya at the redshift of the quasar is on the far right
  of each panel.
  {\it{Bottom panels:}} The smoothed spectra with optical depths
  modified assuming an $R_{\rm eq}$ of the size quoted in
  Table~\ref{tab:results}. The grey regions mark parts of the
  spectra that were masked out automatically.  
  }
  \label{fig:before_after_real}
\end{figure*}

\section{Analysis}

\begin{figure}
  \centering
    \includegraphics[width=8.5cm]{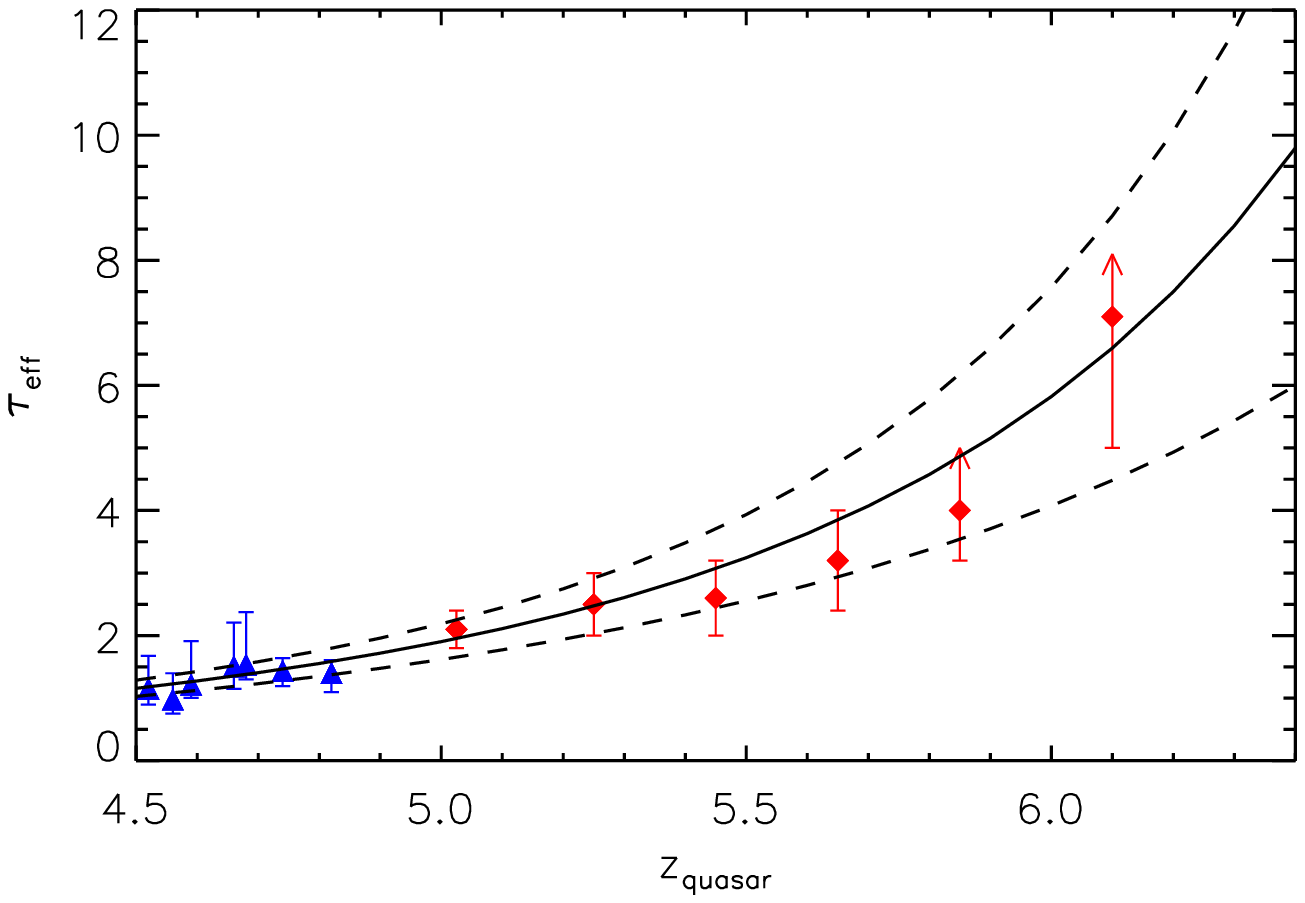}
  \caption{Our adopted fit to the evolution of $\tau_{\rm eff}$
  with redshift. The blue triangles are from measures of the
  flux in the forest from \citet{Songaila2004a}, whilst the red
  diamonds are the binned values from \citet{Fan2006c}. The
  solid line gives the \citet{Becker2007a} relation for the
  evolution of $\tau_{\rm eff}$ with redshift, which we adopt
  for this study. The $1\sigma$ error in $\tau_{\rm eff}$ is
  marked with the dashed lines and was taken to be 10 per cent
  at $z=4$, 15 per cent at $z=5$ and 30 per cent at $z=6$, with
  the error at other redshifts calculated by quadratic extrapolation.
  }
  \label{fig:becker}
\end{figure}

\begin{figure}
  \centering
    \includegraphics[width=8.5cm]{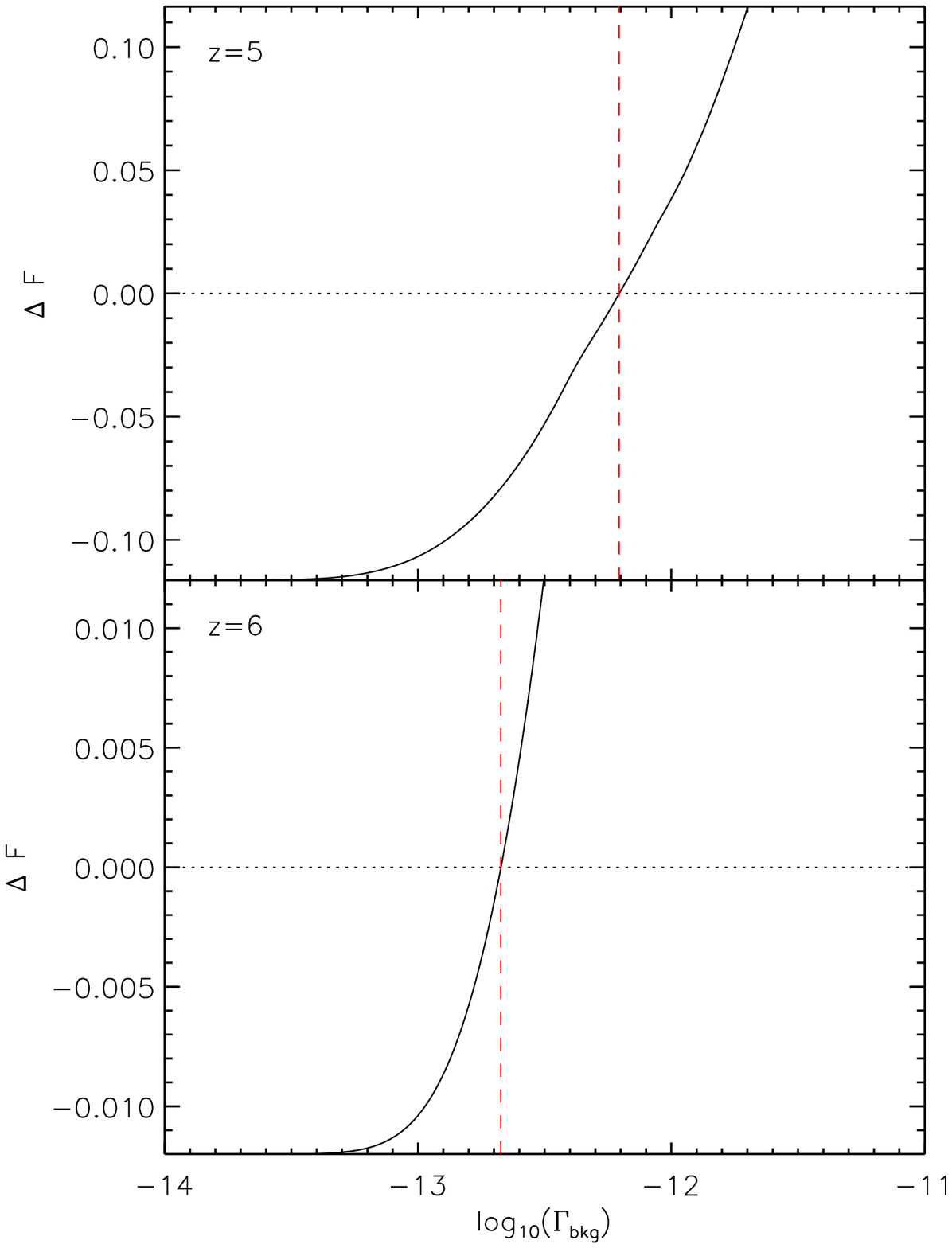}
	\caption{
	{\it {Top panel:}} The variation of $\Delta F$ with
	$\log(\Gamma_{\rm bkg})$ of a simulated quasar spectrum
	at $z=5$. The dotted line marks $\Delta F=0$ and the
	dashed red line	shows the estimated value of
	$\log(\Gamma_{\rm bkg})$. For this particular	input
	spectrum $\Delta F=0$ at $\log(\Gamma_{\rm bkg}) = -12.206$.
	The	curve asymptotes as the mean flux becomes zero.
	{\it {Bottom panel:}} Same as the top but for a simulated
	quasar at $z=6$. The curve is steeper, and so if the mean
	flux has been overestimated by some factor, this will
	result in a smaller change in the estimated
	$\log(\Gamma_{\rm bkg})$ at $z=6$ than at $z=5$.
	}
	\label{fig:max_like_new}
\end{figure}

\subsection{Proximity effect formalism}
\label{sect:details}
As mentioned in Section~\ref{sect:intro}, the ionising
flux from a quasar will locally dominate over the UVB
in setting the ionisation state of the IGM. This leads to
increased transmission near to the quasar compared
to the transmission in the forest (the `proximity effect').
BDO first translated this increased transmission (through
a reduction in the number of strong absorption troughs)
into a measure of the photoionisation rate of hydrogen
caused by the UVB, $\Gamma_{\rm bkg}$. More recently,
measurements have been made of the proximity effect using
flux transmission statistics rather than line-counting
as done by BDO. The average transmission
measured close to the quasar is thereby compared to that
of the average \Lya forest \citep[e.g.][]{Liske2001a}.

The optical depth, $\tau$, is related to the normalised
flux, $F = e^{-\tau}$. In the simplest model where the
quasar lies in a typical region of the IGM, and neglecting
all motion of the gas and any temperature gradients that
may exist as one approaches the quasar, the optical depth
at any point can be described as
\begin{equation}
\tau = \tau_{\rm forest} [1 + \omega(r)]^{-1}~,
\label{tau-mod}
\end{equation}
where $\tau_{\rm forest}$ is the optical depth that
would be measured in the absence of the quasar (i.e.
the typical value for the forest at that redshift), and
\begin{equation}
\omega(r) = \frac{\Gamma_{\rm q} (r)}{\Gamma_{\rm bkg}}~.
\label{omega-def}
\end{equation}
Here, $\Gamma_{\rm q} (r)$ is the \HI photoionisation
rate of the quasar at proper distance $r$, and
$\Gamma_{\rm bkg}$ is the \HI photoionisation rate of the
UVB, assumed to be spatially uniform for a given redshift.
The distance from the quasar, $r$, is approximately
\begin{equation}
r \simeq \frac{c}{H(z)}\frac{\Delta z}{1+z}~.
\label{r-L-def}
\end{equation}
We further define a characteristic length $R_{\rm eq}$
to be the distance from the quasar where the
photoionisation rate from the UVB equals that from the quasar,
i.e. $\Gamma_{\rm q}(R_{\rm eq}) = \Gamma_{\rm bkg}$.
With this definition, Equation~\ref{tau-mod} can then
be rewritten as
\begin{equation}
\tau = \tau_{\rm forest}\left[1+\left(\frac{r}{R_{\rm eq}}\right)^{-2}\right]^{-1}~.
\label{new-tau}
\end{equation}

\begin{figure*}
	\centering
		\includegraphics[width=18cm]{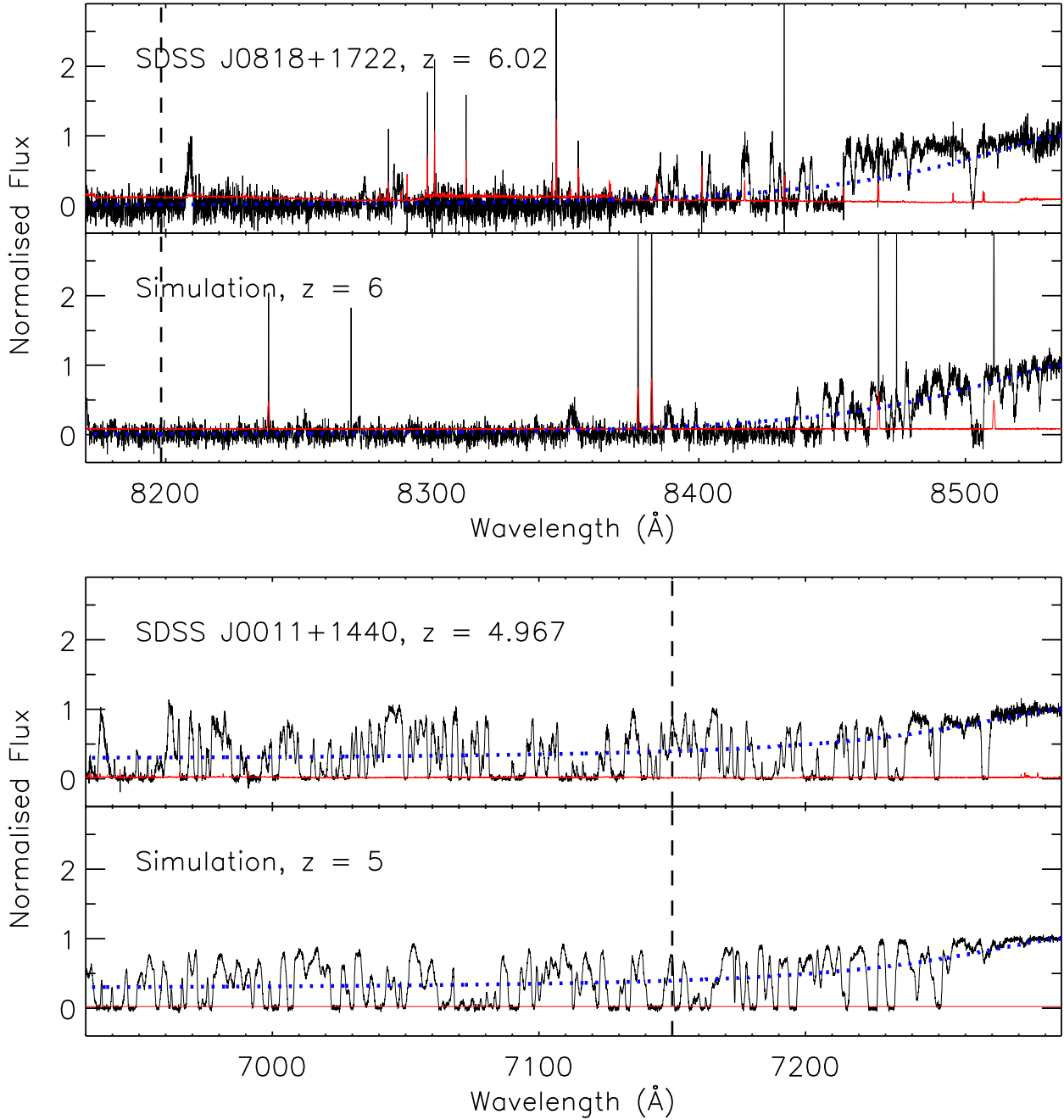}
	\caption{
	{\it {Top:}} The upper panel shows the normalised
	spectrum of SDSS J0818+1722	at $z=6.02$, whilst
	the lower panel is a simulated spectrum at $z=6$
	with the same $R_{\rm eq}$ and $\Gamma_{\rm bkg}$
	as presented in Table~\ref{tab:results}, as well
	as similar noise properties. The red solid line
	is the normalised error spectrum, the black dashed
	line is the derived	value of $R_{\rm eq}$, and the
	dotted blue line is the expected drop off in the
	mean normalised flux. \Lya at the redshift of the
	quasar is on the far right of each plot, and the
	simulated spectrum has been shifted	so that it is
	at the same redshift as the observed spectrum.
	{\it {Bottom:}} The same as above but for SDSS J0011+1440
	at $z=4.967$ and the $z=5$ simulation. Again there
	is a strong similarity between the two spectra. The
	value of $R_{\rm eq}$ is smaller at $z=5$	than at
	$z=6$ because the intensity of the UVB is higher,
	and so the ionising flux from	the quasar dominates
	over the background out to a shorter distance.
	}
	\label{fig:real_vs_fake}
\end{figure*}

It must be emphasised that $R_{\rm eq}$ is different
to proximity region sizes as defined in the literature.
At high redshift the proximity region `size' is 
typically defined to be the maximum extent of the
enhanced transmitted flux
\citep[e.g. the first point at which the transmission
drops to 0.1 in the spectrum when smoothed with a
20~\AA~filter,][]{Fan2006c,Carilli2010a},
but this is an observational, rather than a physical
quantity. In this paper we choose to define the
proximity region size as the scale length out to
which the ionising flux from the quasar dominates
over that from the background, i.e. where $\omega = 1$.
As such any comparison between our proximity region
sizes (values of $R_{\rm eq}$) and those found by
other methods for the same quasars should be made
keeping this difference in mind. For further
discussion on the distribution of proximity region
sizes at high redshift see \citet{Maselli2009a} and
\citet{Bolton2007b}.

\begin{table*}
\centering
  \caption{The contribution to the error and
  systematic shift in $\log(\Gamma_{\rm bkg})$ from
  a variety of causes estimated from 1000 simulated
  spectra at $z=5$ and $z=6$, each with either
  $R_{\rm eq} = 10$ or $5$ proper Mpc. The errors
  from specified properties were found to have a
  nearly Gaussian distribution in $\log(\Gamma_{\rm bkg})$.
  Each property, $x_i$, contributes an error
  $\sigma_{x_i}$ and a systematic shift, $\epsilon_{x_i}$
  (both measured in dex). The two instrumental
  resolutions represent those of MIKE and HIRES at
  $z=5$ and $z=6$ respectively. For the thermal
  proximity effect the temperature of the gas in
  the closest 5 proper Mpc has been raised by $10^4$~K,
  similar to if the quasar had ionised \HeII (see
  Section~\ref{sect:temp}). The total error is
  calculated by adding the individual errors in
  quadrature, whilst the total shift is the sum
  of the individual shifts. For comparison, we also
  simulated spectra with all possible sources of
  error included simultaneously (`model'). The total
  error, as well as the total systematic shift, are
  similar to those expected from combining the
  individual effects.
  }

  \begin{tabular}{c|c|c|c|c|c|c|c|c|c|c|c|c}
    \hline
    \hline
       & \multicolumn{6}{|c|}{$z=5$}                 & \multicolumn{6}{|c|}{$z=6$}                 \\
       & \multicolumn{3}{|c|}{$R_{\rm eq} = 10$~Mpc} & \multicolumn{3}{|c|}{$R_{\rm eq} = 5$~Mpc}
       & \multicolumn{3}{|c|}{$R_{\rm eq} = 10$~Mpc} & \multicolumn{3}{|c|}{$R_{\rm eq} = 5$~Mpc}  \\
    Property                & Value   & $\sigma_{x_i}$ & $\epsilon_{x_i}$ & Value   & $\sigma_{x_i}$ & $\epsilon_{x_i}$
                            & Value   & $\sigma_{x_i}$ & $\epsilon_{x_i}$ & Value   & $\sigma_{x_i}$ & $\epsilon_{x_i}$ \\
  \hline
  
    $\log{L_{\nu_0}}$       & 31.37           & 0.07 & 0.01  & 30.77 & 0.09 & 0.01  & 31.22           & 0.07 & 0.01  & 30.62 & 0.09 & 0.01  \\
    $v_{\rm pec}$           & -               & 0.09 & 0.05  &       & 0.12 & 0.05  & -               & 0.04 & -0.02 &       & 0.07 & 0.00  \\
    Sightline $\delta F$    & 21\%            & 0.13 & 0.02  &       & 0.14 & 0.01  & 43\%            & 0.10 & 0.02  &       & 0.13 & 0.02  \\
    $\Delta \tau_{\rm eff}$ & 15\%            & 0.22 & 0.03  &       & 0.23 & 0.01  & 30\%            & 0.45 & 0.10  &       & 0.59 & 0.11  \\
    $\Delta z$              & 0.005           & 0.18 & -     &       & 0.31 & -     & 0.01            & 0.20 & -     &       & 0.39 & -     \\
    $\Delta L_{\nu_0}$      & 30\%            & 0.14 & -     &       & 0.14 & -     & 30\%            & 0.15 & -     &       & 0.14 & -     \\
    $S/N$                   & 20              & 0.13 & 0.16  &       & 0.11 & 0.29  & 20              & 0.03 & 0.04  &       & 0.09 & 0.09  \\
    Inst. res.              & 13.6\kmps       & 0.03 & 0.05  &       & 0.04 & 0.03  & 6.7\kmps        & 0.02 & 0.01  &       & 0.04 & 0.00  \\
    Halo host mass          & $10^{13} \Msun$ & 0.10 & 0.12  &       & 0.16 & 0.33  & $10^{13} \Msun$ & 0.05 & 0.03  &       & 0.09 & 0.09  \\
    Thermal prox. effect    & $10^4$~K        & 0.07 & -0.10 &       & 0.01 & -0.17 & $10^4$~K        & 0.02 & -0.11 &       & 0.09 & -0.19 \\
  \hline
    Total                   &                 & 0.41 & 0.33  &       & 0.51 & 0.57  &                 & 0.54 & 0.08  &       & 0.76 & 0.13  \\
  \hline
    Model                   &                 & 0.43 & 0.31  &       & 0.56 & 0.57  &                 & 0.54 & 0.09  &       & 0.86 & 0.18  \\
\end{tabular}
\label{tab:err_budget}
\end{table*}

A measurement of the UVB intensity can be
expressed as a value of $R_{\rm eq}$ if the flux
of ionising photons from the quasar and its
fall-off with distance are known. We assume a
$1/r^2$ fall-off, although possible deviations
from this are discussed in Section~\ref{sect:LLS}.
For this the flux at the Lyman limit, $f_{\nu_0}$,
needs to be determined. At the redshifts of the
quasars discussed in this paper insufficient flux
is transmitted through the forest for direct
measurement of $f_{\nu_0}$, and so we extrapolate
the continuum flux (measured at a rest wavelength
of 1280~\AA~as described in Section~\ref{sect:obs_spec})
by assuming a power law relation of the form
$f_\nu \sim \nu^{-\alpha}$. The value of $\alpha$
used in this paper is $1.61 \pm 0.86$ and is
based on the $z <2$ radio-quiet quasar sample of
\citet{Telfer2002a}, who measured this index in
the range 500 to 1200~\AA~for 39 individual AGN.
The error quoted is the RMS scatter of that sample.
By contrast, \citet{Scott2004a} found a harder
index ($\alpha = 0.74$) in their $z<0.67$ sample.
The \citet{Scott2004a} sample, however, covers a
luminosity range that is an order of magnitude
lower than either the \citet{Telfer2002a} sample,
or the quasars analysed here. The \citet{Telfer2002a}
mean value and dispersion should therefore be the
most appropriate for this study. Using $f_{\nu_0}$,
the luminosity of the quasar at the Lyman limit,
$L_{\nu_0}$, is calculated as
\begin{equation}
L_{\nu_0} = 4 \pi d_{L}^{2} \frac{f_{\nu_0}}{(1+z_{\rm q})}~,
\label{L(v_0)-def}
\end{equation}
where $d_{\rm L}$ is the luminosity distance to
the quasar. The quoted errors in $L_{\nu_0}$
take into account the error on $\alpha$, the error
in $d_{\rm L}$ (from the error in the redshift),
and the error in the apparent magnitude of the
continuum, $m_{1280}$. For quasars with $z_{\rm q}<5.5$
the error in $m_{1280}$ is calculated from the
error in the measured continuum flux for SDSS
spectra as described in Section~\ref{sect:obs_spec},
whilst for those with $z>5.5$ it is taken to
be the same as the error on the SDSS $z$-band
photometry. For a given distance from the
quasar, $r$, the Lyman limit flux density is
\begin{equation}
F^{\rm Q}_{\nu_0}(r) = \frac{L_{\nu_0}}{4 \pi r^2}~.
\label{F^Q(v0)-def}
\end{equation}

The photoionisation rate of \HI (in units
of $\rm s^{-1}$) by a source of UV flux is
given by
\begin{equation}
\Gamma = \int^{\infty}_{\nu_0} \frac{4 \pi J(\nu) \sigma_{HI}(\nu)}{h \nu} d \nu~,
\label{big-integral}
\end{equation}
where $J(\nu)$ is the intensity of the source,
$\sigma_{HI}(\nu)$ is the ionisation cross-section
of neutral hydrogen, and $h$ is in this case
Planck's constant. By definition
$J^{\rm bkg}(\nu_0) = J^{\rm Q}(\nu_0)$ at $r = R_{\rm eq}$,
and so using
$\sigma_{HI}(\nu) = 6.3 \times 10^{-18}(\nu_0/\nu)^{2.75}~ \rm cm^2$
\citep[][note that the exponent is often
approximated as 3]{Kirkman2008a} and integrating
Equation~\ref{big-integral} for the photoionisation
rate by the background gives
\begin{equation}
\Gamma_{\rm bkg} = \frac{9.5 \times 10^8 F^{\rm Q}_{\nu_0}(R_{\rm eq})}{(\alpha+2.75)}
\label{final-gamma}
\end{equation}
in units $\rm s^{-1}$, where $F^{\rm Q}_{\nu_0}(R_{\rm eq})$
is the Lyman limit flux density in $\rm erg~cm^{-2}$
evaluated at a distance $R_{\rm eq}$ away from the
quasar. Therefore, using Equation~\ref{F^Q(v0)-def},
$\Gamma_{\rm bkg}$ can be expressed as
\begin{equation}
\Gamma_{\rm bkg} = \frac{9.5 \times 10^8 L_{\nu_0}}{(\alpha+2.75)4 \pi R_{\rm eq}^2}~,
\label{final-final-gamma}
\end{equation}
where $L_{\nu_0}$ is in erg~s$^{-1}$~Hz$^{-1}$ and
$R_{\rm eq}$ is in cm.

\subsection{Measurement method}
\label{sect:application}

The value of $\Gamma_{\rm bkg}$ was inferred for
each quasar by increasing the optical depths
of each pixel by the expected effect of the
quasar, using Equation~\ref{new-tau}, until the
mean flux across the spectra was the same as
that expected from the \Lya forest at that
redshift. Details of this procedure are given
below.

The analysis was carried out on the section of
the spectrum immediately bluewards of the \Lya
line and $40h^{-1}$ comoving Mpc ($\sim 8-9$
proper Mpc, or $\Delta z \sim 0.10-0.15$) in
length. Each pixel is converted from a normalised
flux into an optical depth using $\tau = -\ln(F)$.
For a given quasar luminosity and trial value of
$\Gamma_{\rm bkg}$, Equation~\ref{final-final-gamma}
can be inverted to give a trial value of $R_{\rm eq}$.
Using this trial value of $R_{\rm eq}$, the factor
$[1 + \omega(r)]^{-1}$ can be calculated for every
pixel in the spectrum. The pixel optical depths are
then adjusted, before being converted back into
fluxes. This process is continued until the section
of the spectrum resembles the \Lya forest at the
same redshift. The most robust statistic for
determining this was found to be the difference
between the mean flux in the spectrum and the mean
flux of the \Lya forest, $F_{\rm forest}$. The
difference between the two, $\Delta F$, was defined as
\begin{equation}
\Delta F = \left\langle F_{\rm new} \right\rangle - F_{\rm forest}~,
\label{delta_f-defn}
\end{equation}
where $F_{\rm new} = e^{-\tau_{\rm new}}$ and
$\tau_{\rm new} = \tau[1+\omega(r)]$. 
The value for which $\Delta F =0$ we took as
our estimate of $\Gamma_{\rm bkg}$.
The mean flux, $\left\langle F_{\rm new} \right\rangle$,
was calculated as a weighted mean, with each
pixel weighted by the inverse variance of the
flux. For a given original error in the
transmission, $\sigma _F$, then
$\sigma _{F_{\rm new}} = \sigma _F[1+\omega(r)]^{-1}F_{\rm new}/F$.
The weighted mean flux was then calculated as
$ \left\langle F_{\rm new} \right\rangle = \sum{(F_{\rm new}/\sigma _{F_{\rm new}}^2)}/\sum{(1/\sigma _{F_{\rm new}}^2)}$.
In order to stop $\sigma _{F_{\rm new}}$ from
reaching zero and strongly biasing the weighting,
a noise `floor' was used such that
$\sigma _{F_{\rm new}} \geq 0.01$.

\begin{figure*}
	\centering
		\includegraphics[width=18cm]{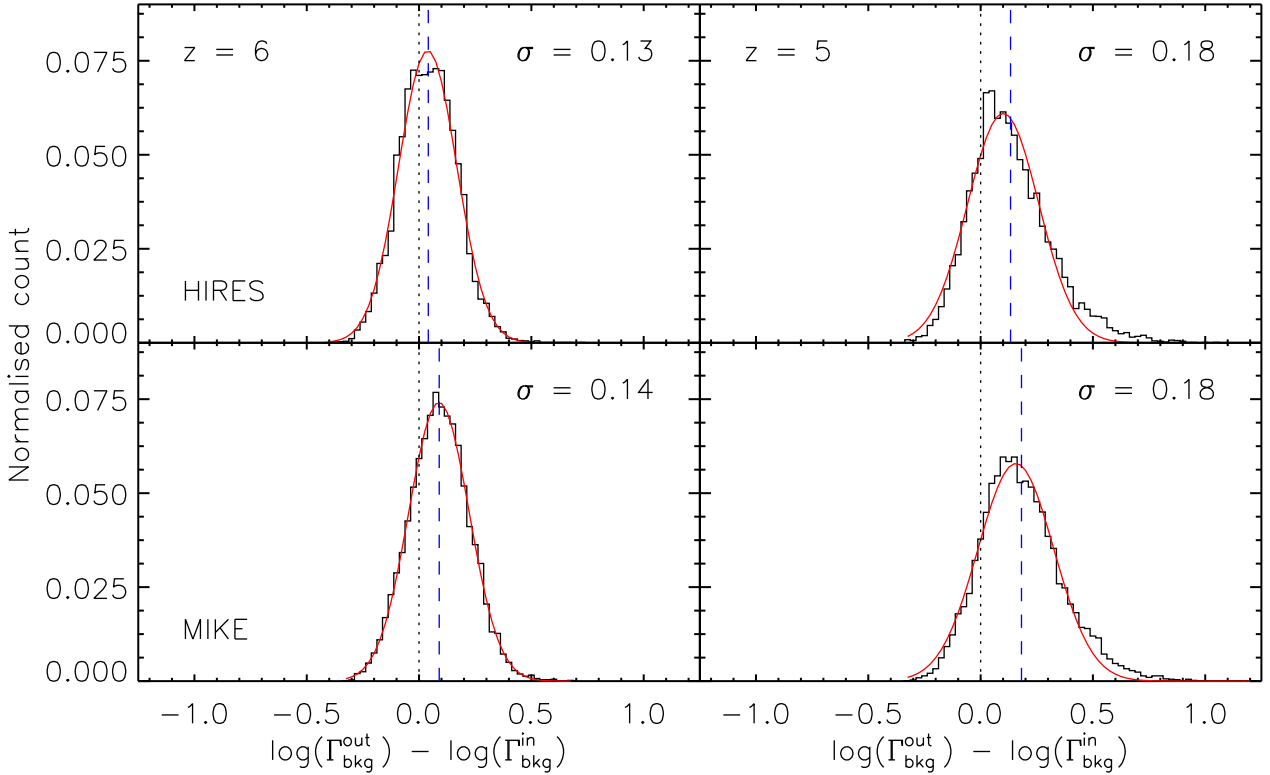}
	\caption{The deviation of the estimated value of
	$\log(\Gamma_{\rm bkg})$ from the input value for $10\,000$
	simulated spectra at $z=6$ (left) and $z=5$ (right) with
	input proximity regions of 10 proper Mpc in size. The
	spectra had a fixed $S/N = 20$, and the resolution of either
	the HIRES (top) or MIKE (bottom) spectrographs. The dotted
	black line marks the input value, whilst the blue dashed
	line marks the mean of the simulated data set. The spread
	is approximately Gaussian in $\log\Gamma$ with a small
	systematic offset caused by noise in the spectra. Each
	offset in the estimated $\Gamma_{\rm bkg}$ corresponds to
	an overestimate of the input value. The width of the
	Gaussian and size of the systematic offset varied with
	redshift, $S/N$, luminosity, and the resolution. Detailed
	modelling of the effect of each of these parameters was
	carried out to calculate the statistical error and bias in the
	estimated $\Gamma_{\rm bkg}$ for both individual quasars
	and a grouped sample. A detailed error budget is presented
	in Table~\ref{tab:err_budget}, and more figures are
	presented in the Appendix.
	}
	\label{fig:check_gamma}
\end{figure*}

\begin{figure*}
	\centering
		\includegraphics[width=18cm]{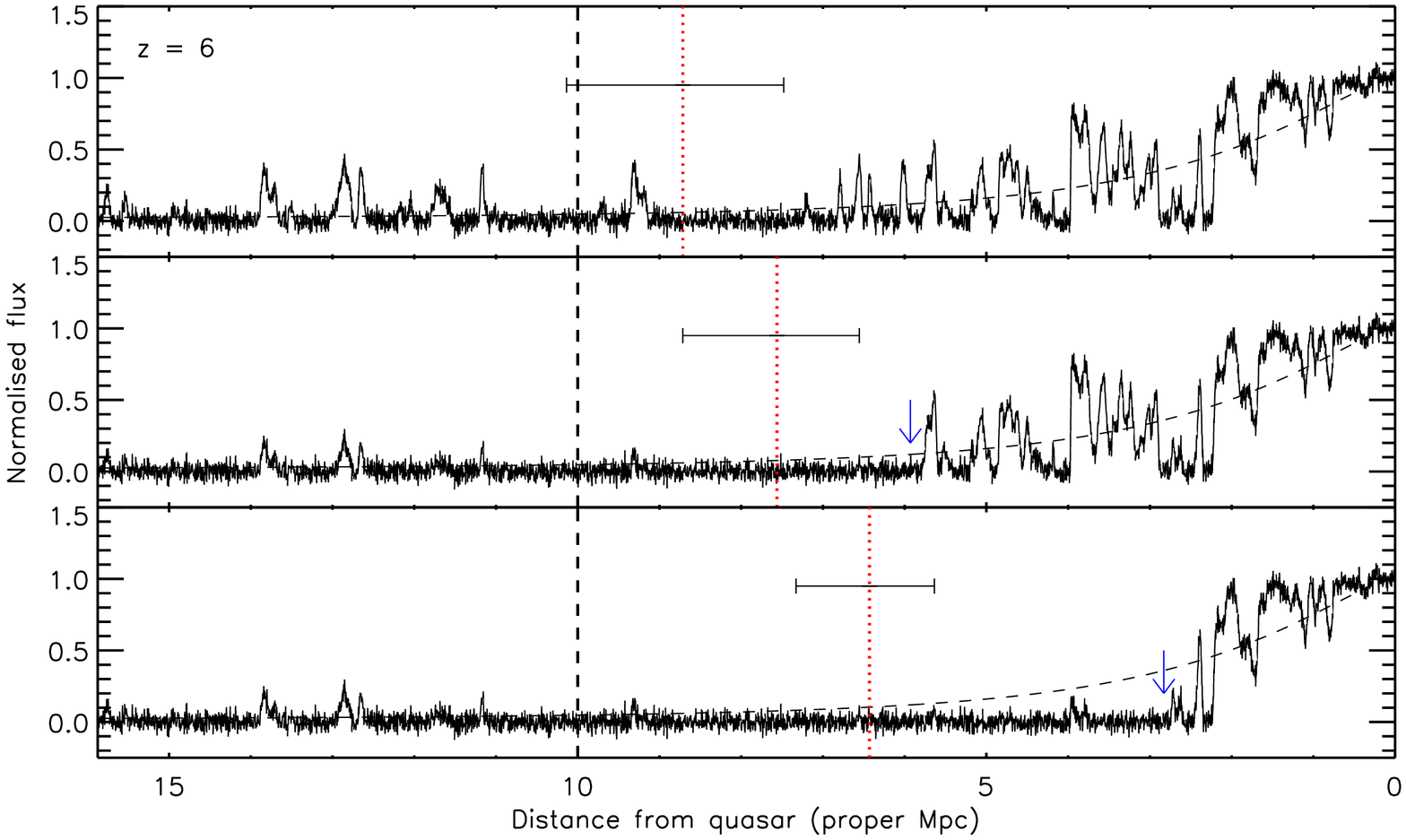}
	\caption{
	Demonstration of the impact of Lyman limit systems
	(see Section~\ref{sect:LLS}).
	{\it {Top:}} A standard simulated spectrum at $z=6$
	with $R_{\rm eq}=10$	proper Mpc and $S/N=20$. The
	thick black dashed line marks this input $R_{\rm eq}$
	whilst the dotted red line shows the one estimated
	by our method. The error bars correspond to	the
	statistical error in the estimated $R_{\rm eq}$,
	roughly 15 per cent. The thin black dashed line is
	the expected drop in normalised flux for the input
	$R_{\rm eq}$. The method has, within error,	recovered
	the	input $R_{\rm eq}$.
	{\it {Middle:}} The same spectrum as above, but with
	a Lyman limit system with corresponding \Lya
	absorption at the location of	the blue arrow. The
	region of enhanced transmitted flux is slightly
	truncated, and so the estimated $R_{\rm eq}$ is
	slightly smaller than in the case of no	Lyman limit
	system, corresponding to a higher estimated value
	of $\Gamma_{\rm bkg}$. This is a well known bias in
	proximity	effect measurements.
	{\it {Bottom:}} The same spectrum as in the top panel,
	but with the Lyman limit system much closer to the
	quasar,	causing substantial shortening of the apparent
	size of the	proximity region. There is again an
	associated decrease	in the estimated $R_{\rm eq}$,
	and thus a higher estimated	$\Gamma_{\rm bkg}$. However,
	the estimated values of $R_{\rm eq}$ are substantially
	more robust than would be expected from the extreme
	shortening of the apparent extent of the enhanced
	transmitted flux.
	}
	\label{fig:shortening}
\end{figure*}

Trial values of $\log(\Gamma_{\rm bkg})$ ranged
from $-14<\log(\Gamma_{\rm bkg})<-11$ and the
iteration was done until a precision of 0.001 in
$\log(\Gamma_{\rm bkg})$ was reached.
Fig.~\ref{fig:before_after} shows the effect of
the result of this iteration on simulated
sightlines at $z=5$ and $z=6$. The `recovered'
\Lya forest (middle panels) is strikingly
similar to the actual \Lya forest in those
sightlines (bottom panels). Two observed spectra
are presented in a similar fashion in
Fig.~\ref{fig:before_after_real}. The bottom
panels show the spectra after smoothing to
mitigate the effect of pixel noise (see
Section~\ref{sect:reliable}) and after the
optical depths have been modified using the
estimated value of $R_{\rm eq}$ presented in
Section~\ref{sect:results}.

The assumed value of $F_{\rm forest}$ is
calculated at each redshift from the fit to
measured optical depths of the \Lya forest
in \citet{Becker2007a}. The fit is reproduced
along with some observed values of $\tau_{\rm eff}$
from the literature in Fig.~\ref{fig:becker}.
It assumes a simple evolution of a lognormal
distribution of optical depths and matches
the observed values excellently from $2 < z < 6$.
Above $z=6$ there are few measurements of the
flux in the forest and so it is unknown if this
relationship still holds. For the $z > 6$ quasars
in this paper it was assumed that it does,
although for large values of effective optical
depth $(\tau > 6)$, a substantial change in
$\tau$ corresponds to only a very small (absolute)
change in $F_{\rm forest}$. Only a small change
in $R_{\rm eq}$, therefore, is required for
$\Delta F$ to again be zero. This is apparent
in Fig.~\ref{fig:max_like_new}, which shows an
example of the results for one of the simulated
spectra at $z=5$ and at $z=6$. If the mean flux
is overestimated by a factor of 5, for example
(corresponding to the dotted line being down at
$-0.10$ and $-0.010$ in the top and bottom panels
respectively) then the shift in the estimated
$\log(\Gamma_{\rm bkg})$ is much greater at $z=5$
($\sim 0.6$ dex) than at $z=6$ ($\sim 0.3$ dex).
The results from the $z > 6$ quasars should
therefore be rather robust to significant
uncertainties in the mean flux.

Probable skylines are identified as regions with
$\sigma _F > 2 \left\langle \sigma _F \right\rangle$,
where the mean is defined over the whole spectrum.
Those pixels, plus 5 pixels either side, are masked
out. The data is also smoothed with a boxcar with
a smoothing window that is proportional to the
amount of noise in the data
(window width $= 10[ \left\langle \sigma _F \right\rangle/0.05]$ pixels),
in order to balance between a smoothly varying
function whilst maintaining some of the contrast
and resolution in the proximity region. Regions
with a smoothed flux below zero were treated as
though they were positive, but remembering the
sign, such that
$F_{\rm new} = -e^{-(-\ln |F|)[1+\omega(r)]}$.
Smoothed fluxes that were greater than or equal
to 1 (i.e. at the continuum) were set to be equal
to 0.99. These effects, designed to limit the
effects of observational artifacts, introduce a
bias such that as the noise increases the method
will systematically overestimate the UVB. These
biases were therefore extensively modelled using
the simulated spectra (see Appendix).

There are several advantages of our method over
other recently presented methods to measure the
intensity of the UVB from the proximity effect.
One of the key attributes of our method is that
the intrinsic distribution of \Lya optical depths
need not be known {\it{a priori}}. This means
that we are not directly dependent on numerical
simulations. Another key advantage is that a slope,
$\beta$, of the column density distribution of
neutral hydrogen is not assumed, as was required
in e.g. \citet{DallAglio2008a}. We believe that
with our method the largest remaining uncertainty
is the evolution of $\tau_{\rm eff}$ at $z>6$.

\subsection{Statistical accuracy of the method}
\label{sect:reliable}
In order to test the accuracy of the method,
simulated spectra were created using random
sightlines through the simulations. The test
spectra were all created using Model C at
$z = 5$ and $6$, and were $40h^{-1}$ (comoving)
Mpc long, constructed by joining together
multiple sightlines. In our standard mock
spectra the proximity region was assumed to
have a size of 10 proper Mpc. This is in
general agreement with the derived values from
the data (see Section~\ref{sect:results}) and
that estimated by \citet{Wyithe2004a}. Assuming
a \citet{Haardt2001a} UVB this corresponds to a
quasar with $M_{1450} = -27.10$ (AB) at $z=6$
and $M_{1450} = -27.47$ (AB) at $z=5$. The
noise was assumed to be Gaussian with a signal-to-noise
($S/N$) of 20, close to the average value of the
$S/N$ in the data.

In order to better mimic the real data, a variety
of artifacts were introduced into the simulated
spectra. Sky line residuals were added with an
associated peak in the error spectrum, as well as
bad pixels for which there was no associated effect
on the error spectrum. Overall the simulated spectra
are qualitatively very similar to the real data, as
demonstrated in Fig.~\ref{fig:real_vs_fake}.

We applied our analysis to artificial spectra
with similar $S/N$ to the data. On average the
method estimated the correct value of
$\Gamma_{\rm bkg}$, with the distribution of
values approximately Gaussian in $\log(\Gamma_{\rm bkg})$.
Several factors were thoroughly investigated
for their effect on the size of the errors,
and any bias on the estimated UVB intensity.
We investigated the effects of peculiar
velocities in the gas, $v_{\rm pec}$, which
will introduce distortions in redshift space to
the expected radial profile of the transmitted
flux. This was done by re-running our analysis
on spectra generated with peculiar velocities
set to zero. Similarly we examined the impact
of sightline-to-sightline variations in the
mean flux away from the global mean flux due
to local fluctuations in the density field.
For this we ran the analysis on spectra from
which the mean flux of that sightline was known
(equivalent to if the quasar was removed perfectly),
and compared it to when simply assuming a global
mean flux. The spread was also dependent on the
noise levels in the simulated spectra, the redshift,
the instrumental resolution, the luminosity, and
the error in $\tau_{\rm forest}$ (which varied with
redshift). This analysis is shown in more detail
in the Appendix. Each contributor to the error
(see Table~\ref{tab:err_budget}) was varied between
reasonable limits for $1000$ simulated spectra at
both $z=5$ and $z=6$, and linearly extrapolated
to other redshifts. This meant that the contribution
to the error budget and associated bias from all
the variables could be calculated for the parameter
space and redshift range present in the data.
Using this analysis an overall statistical error
and bias were determined for each individual quasar.
For demonstration, the dependence on $S/N$ and
instrumental resolution is shown in
Fig.~\ref{fig:check_gamma}. As expected, if the
$S/N$ decreases then the scatter in $\log(\Gamma_{\rm bkg})$
increases. There is also a systematic bias, due
to the smoothing of the spectra. Averaging over
fluxes is not equivalent to averaging over
optical depths, due to the non-linear relation
between the two, and so the more the spectrum
is smoothed the more the optical depths are
underestimated. This effect is strongest for
low fluxes (high optical depths). Consequently,
smoothing (particularly at high redshift)
increases the transmission in the forest so the
fall-off in mean flux is not as steep. This
corresponds to a smaller $R_{\rm eq}$, and so
ultimately $\log(\Gamma_{\rm bkg})$ is overestimated.
Since MIKE has a lower instrumental resolution
than HIRES there is an additional bias brought
in due to a similar effect. Indeed any effect
that smoothes the spectra by averaging over
pixel fluxes will bring in a bias of this nature.
Table~\ref{tab:err_budget} breaks down the sources
of scatter and systematic biases based on $1000$
sightlines that have similar values to those in
the data at $z=5$ and $z=6$. At both redshifts the
errors are dominated by the error in $\tau_{\rm eff}$,
and the errors in redshift are far more important
than the errors in the luminosity. Even though the
error in $\tau_{\rm eff}$ is symmetric (it is
approximated as a Gaussian), the resulting bias is
asymmetric, and becomes more so at high
$\tau_{\rm eff}$ (i.e. high redshifts), as the
equivalent distribution of $F_{\rm forest}$ also
becomes highly asymmetric.

As we noted in Section~\ref{sect:details} we
assume a spatially uniform UVB. However, shortly
after reionization has completed the mean free
path of ionising photons is short enough that
significant spatial variations in the UVB may
exist \citep[e.g][]{Lidz2007a}.
\citet{Mesinger2009a} show however that variations
in the density field at $z \sim 5-6$ dominate over
spatial variation in the UVB, and that even just
after reionization assuming a uniform UVB
underestimates $\Gamma_{\rm bkg}$ by at most a few
percent. We thus do not try to correct for this
uncertain but small effect.

We also performed a joint analysis of several
sightlines simultaneously. These were modelled
in a similar way, with the luminosities, noise
properties and instrumental resolution equal to
that of the component spectra in each bin being
represented. From this a statistical error and
bias could be determined for each bin.

The effect of other sources of systematic bias,
such as Lyman limit systems (LLS, see
Section~\ref{sect:LLS}), were also considered.
These LLS are absorption features that are
optically thick ($\tau > 1$) to Lyman limit
photons, and so prevent the quasar from ionising
as large a volume of the IGM as in their absence.
Consequently they can truncate the extent of
enhanced transmission in a spectrum.
Fig.~\ref{fig:shortening} demonstrates the case
where a Lyman limit system has shortened the
apparent proximity region. As the shortening
becomes more severe, the method underestimates
the true proximity region size (and thus
overestimate the value of $\log(\Gamma_{\rm bkg})$)
as the LLS modifies the assumed $1/r^2$ fall-off
from the quasar. However, our method is more
resilient to this effect than previous proximity
effect determinations of the UVB in the literature.
Consequently it will produce on the whole more
accurate values of UVB, less susceptible to the
effects of Lyman limit systems. This is discussed
in more detail in the following section.

\subsection{Details of potential systematic biases}
\label{sect:syst-bias}
Whilst the proximity effect can be used on both
individual sightlines
\citep[e.g.][]{Williger1994a, Cristiani1995a, Lu1996a}
and in a statistical sense across many spectra
\citep[e.g. BDO;][]{Cooke1997a, Scott2000a}
to help constrain the redshift evolution of
the UVB, several issues must be addressed
before the UVB estimates from the proximity
effect can be considered robust. The proximity
effect is essentially measuring the direct
signature of an enhanced ionisation field near
the quasar by noting a reduced fraction of \HI
extending out across a region of a few physical
Mpc. Therefore any conversion from this
physical size to a value for the UVB will be
highly sensitive to any assumptions made about
the quasar and its environment. There are three
main systematic uncertainties in this conversion.

First, it is assumed that the quasar turned on a
sufficiently long time ago that its proximity
region is in photoionisation equilibrium with the
IGM, and that it has been at its current
luminosity over a similar timescale. The unknown
level of variability in quasar luminosity on the
order of $10^4$ years (the timescales required
for a region of the highly ionised IGM to reach
ionisation equilibrium) means that the measured
proximity region size may not correspond to a
region in ionisation equilibrium, and so any
measurement of the UVB may be biased
\citep{Pentericci2002a, Schirber2004a, DallAglio2008a}.

The second problem is that in order to
calculate an accurate size of the proximity
region, an accurate systemic redshift for
the quasar is needed. Redshifts determined
from broad high-ionisation lines are
underestimates of the systemic redshift
\citep[e.g.][]{Richards2002a}, and an
underestimate in the systemic redshift will
cause the UVB to be overestimated
\citep{Espey1993a}. It is worth noting,
however, that an accurate systemic redshift
can be determined from low ionisation lines,
and so for some spectra this effect can be
avoided. Also it should be noted that at
high redshift the extent of enhanced
transmitted flux will be larger than even
the largest error in systemic redshift present
in the data.

Finally, in the standard proximity effect
analysis by BDO, the IGM within the
proximity region is presumed to have the
same density distribution as the IGM outside
it. Quasars are however hosted by massive
galaxies which are expected to reside in
an environment which has higher than average 
density out to rather large distances  
\citep{Granato2004a, Fontanot2006a, daAngela2008a}.
In that case, the proximity region would
be smaller than expected for a region of
the same size with  an average density
close to the global mean density
\citep{DOdorico2008a}, and thus the UVB
may be overestimated \citep{Loeb1995a}.
This environmental bias should  be worst 
for the most luminous quasars which
presumably lie in some of the most extreme
overdensities at a given redshift
\citep[e.g.][]{Pascarelle2001a, Adelberger2003a, Kim2007a}.
As already discussed this is believed
to be the main reason for the discrepancies
between estimates of the UVB using the
proximity effect and the flux decrement
method
\citep{Rollinde2005a, Guimaraes2007a, Faucher2008a}
at redshift $z \sim 2-4$.

\begin{figure}
	\centering
		\includegraphics[width=8.5cm]{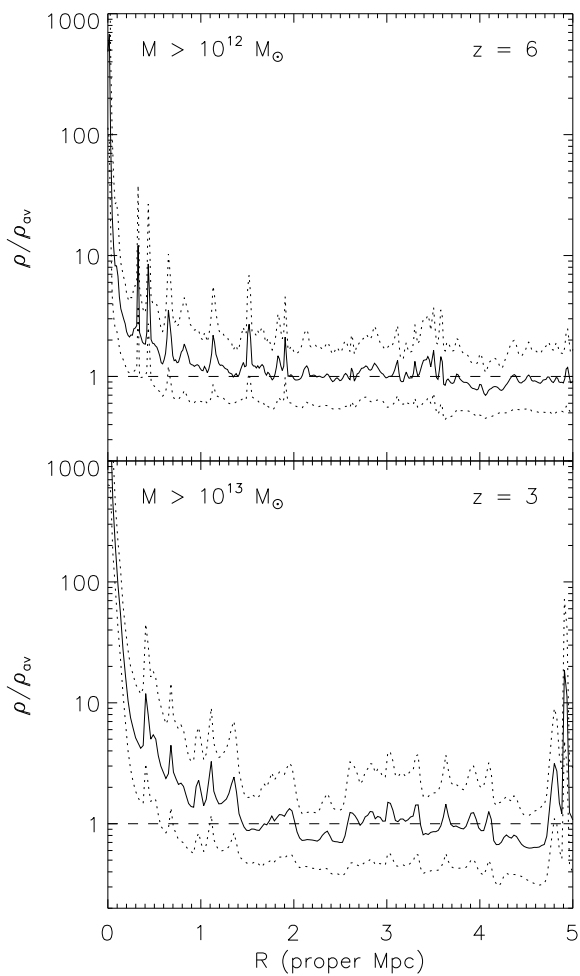}
	\caption{
	Mean density profiles around the most massive
	haloes in our simulation boxes.
	{\it {Top:}}
	The solid line is the average	density field
	in the sightlines that start in haloes of
	masses greater than $10^{12} \Msun$	in Model
	F at $z=6$,	plotted in proper Mpc. The
	dotted lines mark the $\pm 1\sigma$	deviations,
	and the mean density in the simulation
	is marked with the horizontal dashed line.
	Any significant overdensity is restricted
	to a region of $\sim 1$~Mpc in these
	haloes, and the overdense region will be
	even smaller for lower mass haloes.
	{\it {Bottom:}} The same as above but for
	haloes with mass greater than $10^{13} \Msun$
	at $z=3$. These are essentially the same
	haloes as those	at $z=6$, but have become
	more massive through hierarchical growth.
	The overdense region is larger, extending
	out to $\sim 3$~Mpc, with evidence for
	smaller haloes clustered nearby (the small
	peak at $~\sim 5$~Mpc). If the UVB is
	higher at this redshift	then proximity
	region sizes will decrease, and may be
	comparable to the extent of the overdensity.
	In that case measurements of the size of
	the proximity region could depend strongly
	on the mass of the quasar host halo.	
	}
	\label{fig:overdensity_size}
\end{figure}

\begin{figure}
	\centering
		\includegraphics[width=8.5cm]{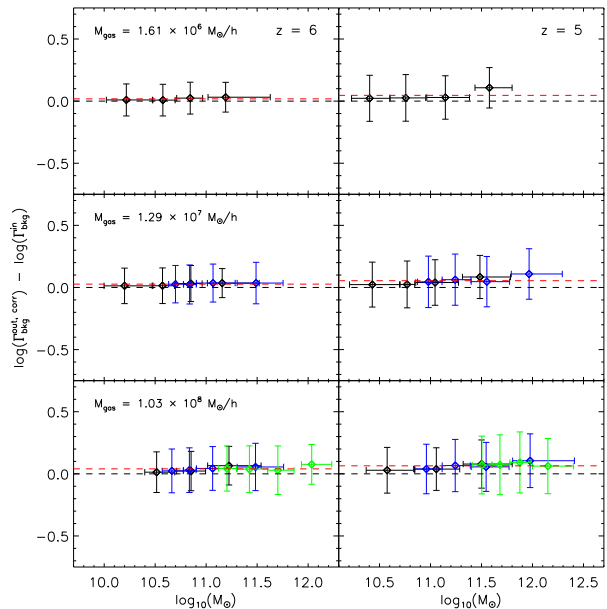}
	\caption{
	{\it {Top row:}} The ratio of the estimated
	$\Gamma_{\rm bkg}$ (corrected for	systematic
	shifts due to noise) to the input
	$\Gamma_{\rm bkg}$ for sightlines starting
	in the 500 most massive haloes in the fiducial
	simulation, Model C. The proximity region in
	each spectra is 10 proper Mpc. On the left
	results are shown for	$z=6$ and on the right
	$z=5$. The black dashed line is for zero bias,
	whilst the red line marks the true average
	value	estimated. For these haloes there is
	a systematic overestimation of $\lesssim 0.1$ dex,
	but apparently no dependence on halo mass.
	{\it {Middle row:}} Same as above, but for
	Models B (black) and E (blue). Model B has
	been resampled to the same pixel size as
	Model E.
	{\it {Bottom panel:}} Same as above but for
	Models A (black), D (blue) and F (green).
	Due to the low resolution and small box size
	in Model A only 100 haloes could be identified
	at $z=6$ and 250 haloes at $z=5$. Models A and
	D have been	resampled to the same pixel size
	as Model F.	Lowering the mass resolution
	introduces a systematic overestimation in
	$\log(\Gamma{\rm bkg})$. This is due to the
	fact that the voids are under-resolved and
	so the proximity regions appear systematically
	smaller. The larger pixel sizes used in the
	lower mass resolution simulations also smooth
	the transmitted flux, so the proximity region
	is again underestimated (see
	Section~\ref{sect:reliable}). This effect is
	the stronger of the two. With these offsets
	taken into account, there is still no
	significant increase in $\Gamma_{\rm bkg}$
  with halo mass, even in the bottom panel,
  which includes an $80h^{-1}$ comoving Mpc
  box with haloes of mass $>10^{12} \Msun$.
	}
	\label{fig:haloes}
\end{figure}

Another environmental effect that is often
neglected is the consequence of the quasar
heating the surrounding IGM via ionisation
of \HeII. \citet{Bolton2010a} found that the
gas within $\sim 5$ proper Mpc of the $z \approx 6$
quasar SDSS J0818+1722 was $\sim 10^4$~K
hotter than that presumed for the general
IGM at that redshift. Higher temperatures
within the proximity region will lead to
more transmission, and so the UVB might
then be underestimated.

All of these effects were investigated
with the simulations at $z=5$ and $z=6$
to try and quantify how important they
might be in the real data, and so each
of these points will now be discussed
in detail.

\subsubsection{Effect of luminosity and redshift errors}
\label{sect:lum-z}
Investigation with both the real and simulated
spectra showed that the measured error on the
UVB is dominated by errors in the redshift
and the effective optical depth of the forest,
and not by errors in the quasar luminosity.
Even random errors in luminosity of up to 40
per cent had only a comparable effect to
small errors ($\sim 0.01$) in redshift. 
Of course, if the variability in quasar
luminosity is not simply an extra source of
random scatter, and in fact is instead
systematic (for example, if the quasars are
only observed in their brightest stages) then
there will  be a systematic bias as well as
an additional random error. The method ultimately
is sensitive to the gradient of the fall-off
in the mean transmitted flux throughout the
spectrum, and so will give the same estimate
for $R_{\rm eq}$ regardless of the quasar
luminosity. Thus, if all the quasars are
radiating at a systematically brighter
luminosity than the one that established their
proximity region, then the values of
$\Gamma_{\rm bkg}$ will have been overestimated
by the same amount, since they are linearly
related in Equation~\ref{final-final-gamma}.

Several of the quasars in this paper have a very
accurate systemic redshift from either CO or
\MgII emission lines \citep{Carilli2010a}, and
the high resolution spectra themselves allow
for a reasonably accurate determination of the
onset of the \Lya forest. Consequently almost
all the redshift errors are $\leq 0.01$,
corresponding to a size much smaller than the
proximity region. We therefore believe that
these two possible sources of error have been
suitably dealt with.

\subsubsection{Effect of quasars lying in overdensities}
\label{sect:halo-mass}

One of the more contentious assumptions used in
proximity region measurements is that the
environment of the quasar within the proximity
region is similar to that of the general IGM.
The quasar should lie in an overdense environment,
and therefore could struggle to ionise as large
a volume as it would in the general IGM, thus
reducing the apparent size of the proximity region.
As discussed already there have recently been a
number of studies suggesting that proximity effect
measurements may indeed suffer from a substantial 
bias due to this
\citep{Rollinde2005a,Guimaraes2007a,Faucher2008a}.

We have therefore tried to estimate how much such
a bias may affect our measurements, which are at
significantly higher redshift than previous studies.
As a first attempt we looked into the effect of
placing the quasar at the centre of massive dark
matter haloes in our fiducial simulation and
simulations with different box size and mass
resolution.

\begin{table*}
\centering
  \caption{Summary of the parameters of the
  different power laws used to investigate the
  effect of LLS on both individual and grouped
  spectra. The number of expected LLS in a
  random sightline through Model C is calculated
  using Equation~\ref{n_lls} with $z_{\rm max}$
  equal to the simulation redshift and
  $z_{\rm max}-z_{\rm min}$ corresponds to
  $20h^{-1}$ comoving Mpc. SC10 parameterise
  their power law with $N_0$ evaluated at
  $z=3.5$ such that instead of
  $N(z) = N_0(1+z)^{\gamma}$ they have
  $N(z) = N_{3.5}[(1+z)/4.5]^{\gamma}$. The
  shifts, $\epsilon$, quoted are the average
  overestimation in $\log(\Gamma_{\rm bkg})$
  due to the presence of LLS. For the
  individual spectra, those at $z=6$ have
  HIRES resolution, and those at $z=5$ have
  MIKE resolution. For the grouped spectra
  the shifts correspond to the $z \sim 5$
  and $z \sim 6$ redshift bins, as described
  in Section~\ref{sect:results}. The final
  row represents the results based upon a
  power law consistent with the 1 $\sigma$
  upper limit of the SC10 parameters.
  }

  \begin{tabular}{c|c|c|c|c|c|c|c|c|c|c}
    \hline
    \hline
     &                              &                         &                                  &                                         & \multicolumn{3}{|c|}{$z=5$}           & \multicolumn{3}{|c|}{$z=6$}       \\
     & Redshift range               & $N_0$                   & $N_{3.5}$                        & $\gamma$                                & $n_{\rm lls}$ & $\epsilon$ (ind) & $\epsilon$ (bin) & $n_{\rm lls}$ & $\epsilon$ (ind) & $\epsilon$ (bin) \\
  \hline
SL94 & $0.40 < z < 4.69$            & $0.27^{+0.20}_{-0.13}$  & --                               & $1.55 \pm 0.45$                         & 0.2169    & 0.15        & 0.09        & 0.3459    & 0.17        & 0.08 \\
P03  & $2.40 < z < 4.93$            & $0.07^{+0.13}_{-0.04}$  & --                               & $2.45^{+0.75}_{-0.65}$                  & 0.2810    & 0.20        & 0.12        & 0.5146    & 0.26        & 0.11 \\
SC10 & \multirow{2}{*}{$0 < z < 6$} & \multirow{2}{*}{$0.15^a$} & \multirow{2}{*}{$2.80 \pm 0.33$} & \multirow{2}{*}{$1.94^{+0.36}_{-0.32}$} & 0.2441     & 0.17        & 0.10        & 0.4134    & 0.22        & 0.09 \\
SC10 $1\sigma$  &                   &                         &                                  &                                         & 0.3022    & 0.21        & 0.12        & 0.5409    & 0.27        & 0.12 \\
  \hline
     & \multicolumn{10}{|l|}{$^a$ Derived from their $N_{3.5}$ and $\gamma$ values.} \\
\end{tabular}
\label{tab:lls}
\end{table*}

For this we have chosen sightlines through
the largest (by total mass) 500 haloes in
Models B-F in 3 perpendicular directions
through the box. Due to the low resolution
and small box size of Model A, only 100
haloes could be identified at $z=6$ and 250
at $z=5$. For each halo the sightline was
adjusted so that it began at the centre of
a massive halo. Since the simulation box
is cyclic the sightlines needed to be
split in half so the sightline only passed
through the overdensity once, and
consequently for each halo 6 sightlines
could be drawn (i.e. leaving the halo from
both sides in 3 perpendicular directions).
The halo sightlines were then spliced with
random sightlines so that the total length
of each spectrum was $40h^{-1}$ comoving
Mpc. The proximity region size was chosen
to be 10 proper Mpc, as in previous tests
with the simulations. The results for the
haloes in the fiducial simulation, Model C,
are shown in the top panel of Fig.~\ref{fig:haloes}.
There is no significant change as the mass
increases, with a Gaussian spread in
$\log(\Gamma_{\rm bkg})$ in each bin of
similar width to that from the random
sightlines (Fig.~\ref{fig:check_gamma}).

The box size for Model C is, however,
only $20h^{-1}$ comoving Mpc, which limits
the maximum halo mass to just over
$4 \times 10^{11} \Msun$. Luminous quasars
are likely to reside in host hales more
massive than this. The best observational
constraints are inferred from the
clustering analysis of quasars combined
with predictions from \LCDM simulations
and suggest host halo masses in the range
$10^{12}-10^{13} \Msun$ \citep[e.g.][]{Bonoli2009a}.
It is important to note here that the
very luminous quasars often used for
proximity effect measurements are more
luminous than those available for
clustering analysis. It is thus perhaps
not surprising that the proximity effect
studies studies by \citet{Rollinde2005a}
and \citet{Guimaraes2007a} suggest that
at $z\sim 2.5-4$ the host haloes of these
very luminous quasars are even more massive.
Fortunately, at the redshifts we consider
here haloes more massive than $10^{13} \Msun$
are not yet expected to have formed in
large enough numbers to be credible
candidates to host even the most luminous
observed quasars \citep{Springel2005b,Sijacki2009a}.
In order to probe overdensities around
haloes in this mass range larger boxes
are needed than we have considered so
far, and so simulations with $40h^{-1}$
and $80h^{-1}$ comoving Mpc box size
were also investigated. These were then
compared with $20h^{-1}$ Mpc boxes with
equal mass resolution. Model B was compared
to Model E (see middle panel in
Fig.~\ref{fig:haloes}), and Model A to
Model D and Model F (bottom panel).
Sightlines through Models A, B and D were
resampled so that the pixel size (and mean
forest flux) was the same as the largest
simulation to which they were being
compared. The highest mass bin in Model F
($>10^{12} \Msun$) still shows no more
significant bias (in comparison to less
massive haloes in Models A and D), seeming
to confirm the results from Model C that,
at least at $z=5-6$, the mass dependence
of the bias is rather weak.

The largest halo mass in the $z=6$
simulation was $10^{12.2} \Msun$, which
may still be nearly an order of magnitude
less massive than the haloes the luminous
quasars used in our study may reside in.
To get a rough estimate of the effect of
more massive haloes we have taken the
density profile of haloes with
$M>10^{13} \Msun$ at $z=3$ in our simulation
(the bottom panel in Fig.~\ref{fig:overdensity_size})
and modified the optical depths of spectra
created from random sightlines in the
$z=6$ simulation by using
$\tau \propto \Delta^{2(\gamma-1)}$,
where $\gamma$ is from the temperature-density
relation, $T(\Delta) \propto \Delta^{\gamma-1}$,
and is taken to be $1.3$. The extra
systematic bias from such a massive halo
is shown in Table~\ref{tab:err_budget}.
The effect is small at $z=6$, and even
though it is more important at $z=5$,
it is still not the main source of bias
or main contributor to the error budget.
Obviously this is just an approximation
of how a halo this massive would affect
the results, as the density profile was
applied to the spectra artificially and
will not be self-consistent with the
other simulation outputs along the
line-of-sight. However, since the effect
is relatively minor we would not expect
it to be a dominant source of error even
if it were modelled self-consistently
using a halo found in a much bigger
simulation box. Since the quasars at
$z=6$ will probably lie in haloes with
$M<10^{13} \Msun$ our estimate should
be an upper limit of the effect of
overdensities, and thus we conclude
that at $z=5-6$ the expected overdensity 
around our quasars is not critically
important.

The most likely explanation for the lack of
dependence of the bias of the UVB estimate
on the mass of the host halo is that the
proximity region is large in comparison
to the size of the overdensity. This is
demonstrated in Fig.~\ref{fig:overdensity_size}
where the overdensity profile is shown at
$z=6$ and $z=3$ for the largest haloes in
Model F ($M>10^{12} \Msun$ and $M>10^{13} \Msun$
respectively). The gas will thereby tend
to be infalling and the corresponding
peculiar velocities will further decrease
the apparent size of the proximity region
in velocity space \citep{Faucher2008a}.
In our simulated spectra the proximity
regions were 10 proper Mpc, which is nearly
an order of magnitude larger than the
overdensity expected in the largest haloes
at $z=6$, and so the properties of the IGM
within the proximity region should not be
strongly biased due to the overdensity.

As a final point, we note that some
clustering of galaxies and faint Active
Galactic Nuclei (AGN) is expected around
luminous quasars
\citep[e.g.][]{Utsumi2010a}. 
The intensity of the UVB in the vicinity
of luminous quasars could therefore be
enhanced due to other nearby sources.
Our knowledge about the clustering of
faint AGN and galaxies around bright
high-redshift quasars is rather sparse.
We can nevertheless use the magnitudes
and positions of the Lyman Break Galaxies
(LBGs) found by \citet{Utsumi2010a} for
a rough estimate. In a field containing
a quasar at $z=6.43$ they found 7 LBGs at
$z > 6.4$, based on their colours. Let us
then calculate $L_{\nu_0}$ for each LBG
from their $z_R$ band magnitude, assuming
they radiate isotropically, are are all
at the same redshift as the quasar, and
have a ratio of far-UV to extreme-UV flux,
$f_{1500}/f_{900} = 22$ \citep{Shapley2006a},
as measured for $z \sim 3$ LBGs. With
projected distances of $2-5$~Mpc the LBGs
would all lie within the typical proximity
zones of our sample if they were indeed at
the same redshift. Even if the LBGs would
have a Lyman continuum as blue as we assume
for the quasars (i.e. $\alpha = 1.61$)
they would increase the size of the proximity
region by as little as $\sim 0.1$ Mpc, and
the UVB would typically be underestimated
by 0.01 dex (3 per cent). There could
obviously be a substantial contribution
from more numerous fainter objects below
the detection limit. The faint end slope
of the luminosity function of LBGs at
$z \sim 6.4$ is not well constrained, but
if we assume a slope of $-1.73$ \citep{Bouwens2010a}
down to zero luminosity in the volume of
the proximity region the UVB is still only
underestimated by 0.04 dex (11 per cent).
This is, however, likely to be an upper limit,
as in reality the 7 LBGs do not have precise
redshifts and may not be within the proximity
region of the quasar. This should outweigh
the possibility that the galaxies may have
a larger value of $f_{1500}/f_{900}$ than
we have assumed here. We therefore conclude
that the effect of nearby LBGs could be
noticeable but appears likely to be small.
Considering the large uncertainties in our
rough estimate we did not try to correct for this.

\subsubsection{Effect of Lyman limit systems}
\label{sect:LLS}
Lyman limit systems (LLS) are due to regions
of neutral gas that are optically thick
($\tau > 1$) to Lyman limit photons
($\lambda_{\rm rest} = 912$~\AA), which
corresponds to a hydrogen column density of
$N(\HI) > 1.6 \times 10^{17}$ atom \psqcm.
Should a LLS lie in the observed line-of-sight
of the quasar then the apparent size of the
proximity region will be shortened, as their
abundance limits the mean free path of ionising
photons
\citep[e.g.][]{Storrie1994a, Miralda2003a, Peroux2003a, Furlanetto2009a}.

\begin{figure*}
	\centering
		\includegraphics[width=18cm]{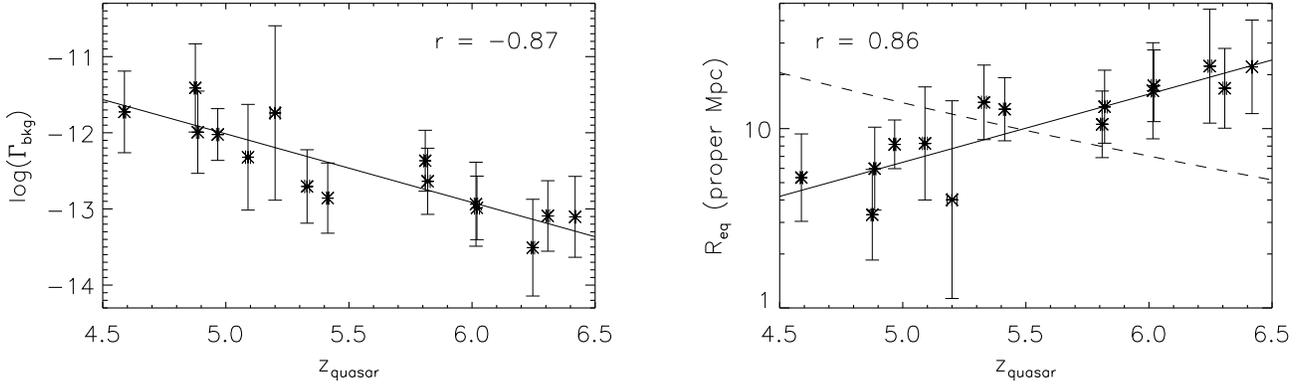}
	\caption{
	{\it {Left panel:}} The bias corrected
	estimates of $\log(\Gamma_{\rm bkg})$ for
	the 15 individual quasars studied in this paper.
	The errors are calculated as the standard
	deviation in $\log(\Gamma_{\rm bkg})$ from
	$10\,000$ Monte Carlo realisations (varying
	the systemic redshift, luminosity and forest
	$\tau_{\rm eff}$)	added in quadrature to the
	statistical error, which was individually
	calculated for each quasar sightline. Biases
	similar to those listed in Table~\ref{tab:err_budget}
	have also been removed. A smooth decline in
	$\Gamma_{\rm bkg}$ with redshift appears
	over this redshift range, with a formal
	correlation coefficient of $-0.87$.
	{\it {Right panel:}} The inferred values
	of $R_{\rm eq}$ for each quasar. The error
	bars were calculated in the same way as
	those for $\Gamma_{\rm bkg}$. There is a
	strong increase towards higher redshifts,
	driven by the declining	intensity of the
	UVB. A fit to this increase is shown by
	the solid line,	and the formal correlation
	coefficient is $0.86$. The dashed	line
	marks the evolution in the mean free path
	of ionising photons from the formula in
	\citet{Songaila2010a}. The size	of the
	proximity region is larger than the mean
	free path at $z \gtrsim 5.5$. The
	implications of this are discussed in
	Section~\ref{sect:comp}.
	}
	\label{fig:results}
\end{figure*}

Studies counting the number of LLS in spectra
struggle at high redshift as, due to the ever
increasing blackness of the forest, features with
$\tau > 1$ are difficult to detect. Consequently there
are very few studies that provide measured LLS
frequency at the redshifts covered in this paper.
LLS can lead to dramatic shortening of the region
of enhanced transmission (see Fig.~\ref{fig:shortening})
and so if abundant at high redshift then they
could be a substantial source of systematic errors.

We inserted LLS into our simulations following the
method presented in Appendix D of \citet{Bolton2007b}.
A density threshold, $\rho_{\rm thresh}$, was chosen
such that the average number of regions in a spectrum 
with $\rho \geq \rho_{\rm thresh}$ was the same as a
given number of LLS expected per sightline. The
neutral fraction within those regions was then set
to unity (i.e. they become self-shielded). As such,
they absorb all the ionising flux of the quasar so
further out in the spectrum the transmission in the
forest is from the UVB alone. The expected number
of LLS in a particular sightline, $n_{\rm lls}$, was
calculated from the power laws presented by
\citet{Storrie1994a}, \citet{Peroux2003a}, and \citet{Songaila2010a},
(hereafter SL94, P03 and SC10 respectively),
where $N(z) \equiv dN/dz = N_0(1+z)^{\gamma}$.
SL94 used quasars covering the range $0.40<z<4.69$, and
suggest that for a sightline covering the redshift
range $z_{\rm max}$ to $z_{\rm min}$ that
\begin{equation}
n_{\rm lls} = \int^{z_{\rm max}}_{z_{\rm min}}N_0(1+z)^{\gamma}~,
\label{n_lls}
\end{equation}
where $N_0 = 0.27$ and $\gamma = 1.55$. For a
particular random sightline in Model C
$z_{\rm max}$ was taken to be the redshift of
the simulation and $z_{\rm max}-z_{\rm min}$
corresponds to the simulation box size. This meant
that $n_{\rm lls} = (0.2169,0.3459)$ at redshift
$z=(5,6)$ respectively. From this (using only the
random sightlines) the threshold densities
were derived as
$\log(\Delta_{\rm thresh}) = (1.6310,1.4080)$
(where $\Delta_{\rm thresh} \equiv \rho_{\rm thresh}/\rho_{\rm av}$)
and those regions were then presumed to be
self-shielded. Similar calculations were
also done for the power laws of P03 and SC10,
and a worst case scenario was investigated
using the $1\sigma$ upper limits of SC10.
The expected overestimations in
$\log(\Gamma_{\rm bkg})$ for each of these
power laws are presented in Table~\ref{tab:lls}
for both individual spectra, and grouped spectra.
Consequently, assuming the predictions of SC10 to
be the most reliable as this is the only study 
probing the redshift range we are interested in,
the effect would be to shift the results of the
individual quasars down by 0.17 (0.22) dex at
$z = 5~(6)$, and possibly up to 0.21 (0.27) dex.

SC10 is the only study into the actual spatial
frequency of LLS at $z = 6$. In both SL94 and P03
they found that towards higher redshifts, $N(z)$
starts to evolve rapidly, and so extrapolations
become much more uncertain. Indeed \citet{Prochaska2010a}
note that both of those papers are subject to biases
that they claim have not been adequately compensated
for, and so they may have overestimated $N(z)$ by up
to a factor of 3. Clearly if $n_{\rm lls}$ is a factor of
3 smaller then the effect on $\Gamma_{\rm bkg}$ would
be greatly reduced. SC10 themselves also warn against
extrapolating to $z>6$ due to the potential rapid change
in the mean free path of ionising photons if reionization
is being probed. Taking all these factors into account,
and treating SC10 as the most reliable power law, we
cautiously estimate using their $1\sigma$ upper limits
on $n_{\rm lls}$ that the binned data points in
Fig.~\ref{fig:z-gamma-others-binned} are possibly
overestimated by $\leq 0.12$ dex.

\subsubsection{Effect of quasars heating the IGM in the proximity region}
\label{sect:temp}
We have implicitly assumed that the temperature
of the quasar proximity region is comparable to
that of the general IGM. This may be not true,
particularly prior to \HeII reionization, when
the ionisation of \HeII by the quasar may heat
the local IGM \citep[e.g][]{Bolton2010a}.
This temperature gradient will cause higher mean
flux in the proximity region, and so the proximity
region size will be {\it{overestimated}} and the
UVB {\it{underestimated}} (i.e. opposite to all
the other environmental biases). We constructed a
toy model in the simulations with the change in
temperature as a step function, such that the gas
within 5 proper Mpc of the quasar was $10^4$~K
hotter than that in the general IGM. This is
similar to the result of \citet{Bolton2010a},
assuming general IGM temperatures at $z=5-6$ of
$\sim 10^4$~K \citep{Becker2010a}. In individual
sightlines this caused $\log(\Gamma_{\rm bkg})$ to
be underestimated by $0.10~(0.11)$ dex at $z=5~(6)$,
and in the binned data by $0.14~(0.21)$.
Consequently, we find that the hotter temperatures
of gas close to the quasar could cause underestimation
in $\log(\Gamma_{\rm bkg})$ comparable to the
overestimations from other environmental effects at
the redshifts of our study.

\begin{figure}
	\centering
		\includegraphics[width=8.5cm]{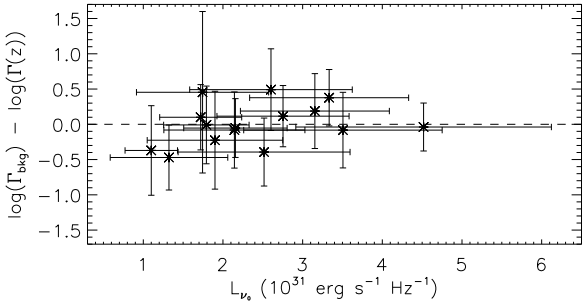}
	\caption{
	Plot of the deviance from the linear fit to
	$\log(\Gamma_{\rm bkg})$ in the left panel of
	Fig.~\ref{fig:results} against the quasar
	luminosity.	The dashed line marks zero
	deviation. All of the points are consistent
	with this line, which suggests that the measured
	values of $\log(\Gamma_{\rm bkg})$ have been
	adequately corrected for systematic biases
	from quasar luminosities.
  }
	\label{fig:lum_ind}
\end{figure}

\section{Results and Discussion}

\begin{table*}
\centering
\begin{minipage}{180mm}
\begin{center}

  \caption{Tabulated results for each of the
  investigated quasar sightlines. Those with
  $\sigma _z$ of 0.002 or 0.007 have systemic
  redshift taken from the CO or \MgII emission
  lines \citep{Carilli2010a} while the others
  use either the \Lya + \NV emission line
  redshift, or (in most cases) use the redshift
  at which the \Lya forest appears to begin.
  Due to the high-resolution of the spectra,
  errors for these are $\leq 0.01$. The
  absolute magnitudes at rest-frame 1450~\AA~
  were calculated from either the published
  continuum magnitudes
  \citep{Fan2001a, Fan2003a, Fan2004a, Fan2006a}
  or the measured fluxes at rest-frame 1280~\AA,
  extrapolated to 1450~\AA~by assuming a power
  law continuum of the form $f_\nu \propto \nu^{-0.5}$.
  }
  \begin{tabular}{c|c|c|c|c|c|c|c|c|c}
    \hline
    \hline
Name            & $z_{\rm q}$ & $\sigma_z$ & $z$ source  & $L^{\rm Q}_{\nu_0}$ & $M_{1450}$ & $R_{\rm eq}$           & $\Gamma_{\rm bkg}$      & $\log(\Gamma_{\rm bkg})$ \\
                &             &            &             & $10^{31}\ergpspHz$  & (AB)       & (Mpc)                  & $10^{-12}\ps$           &                          \\
    \hline
SDSS J1148+5251	& 6.4189      & 0.002      & CO          & $3.154 \pm	0.935$   & --27.81    &	$22.1^{+18.2}_{-10.0}$ & $0.08^{+0.19} _{-0.06}$ &	$-13.10 \pm 0.53$ \\
SDSS J1030+0524	& 6.308	      & 0.007      & \MgII       & $1.721	\pm 0.516$   & --27.15	  &	$16.8^{+11.2}_{-6.7} $ & $0.08^{+0.15} _{-0.05}$ &	$-13.09 \pm 0.46$ \\
SDSS J1623+3112	& 6.247       & 0.007      & \MgII       & $1.100	\pm 0.329$   & --26.67	  &	$22.3^{+24.0}_{-11.6}$ & $0.03^{+0.10} _{-0.02}$ &	$-13.51 \pm 0.64$ \\
SDSS J0818+1722	& 6.02        & 0.01	     & \Lya forest & $2.156	\pm 0.649$   & --27.40	  &	$17.3^{+10.1}_{-6.4} $ & $0.10^{+0.17} _{-0.06}$ &	$-12.99 \pm 0.42$ \\
SDSS J1306+0356	& 6.016	      & 0.007      & \MgII       & $1.792	\pm 0.536$   & --27.20	  &	$16.3^{+13.9}_{-7.5} $ & $0.12^{+0.30} _{-0.08}$ &	$-12.94 \pm 0.55$ \\
SDSS J0002+2550	& 5.82	      & 0.01	     & \Lya forest & $2.754	\pm 0.829$   & --27.67	  &	$13.3^{+7.9} _{-5.0} $ & $0.23^{+0.40} _{-0.15}$ &	$-12.64 \pm 0.43$ \\
SDSS J0836+0054	& 5.810	      & 0.007      & \MgII       & $3.333	\pm 0.998$   & --27.87	  &	$10.6^{+5.6} _{-3.7} $ & $0.43^{+0.65} _{-0.26}$ &	$-12.37 \pm 0.40$ \\
SDSS J0231-0728	& 5.41	      & 0.01       & \Lya forest & $1.323	\pm 0.738$   & --26.87	  &	$12.8^{+6.4} _{-4.3} $ & $0.14^{+0.26} _{-0.09}$ &	$-12.86 \pm 0.46$ \\
SDSS J1659+2709	& 5.33	      & 0.01	     & \Lya forest & $2.516	\pm 1.079$   & --27.57	  &	$14.0^{+8.6} _{-5.3} $ & $0.20^{+0.40} _{-0.13}$ &	$-12.71 \pm 0.48$ \\
SDSS J0915+4924	& 5.20        & 0.02	     & \Lya + \NV  & $1.744	\pm 0.830$   & --27.17	  &	$4.0 ^{+10.3}_{-2.9} $ & $1.82^{+23.60}_{-1.69}$ &	$-11.74 \pm 1.15$ \\
SDSS J1204-0021	& 5.09	      & 0.01	     & \Lya forest & $1.900	\pm 0.851$   & --27.26	  &	$8.3 ^{+8.8} _{-4.3} $ & $0.48^{+1.88} _{-0.38}$ &	$-12.32 \pm 0.69$ \\
SDSS J0011+1440	& 4.967	      & 0.005      & \Lya forest & $4.518	\pm 1.604$   & --28.20	  &	$8.2 ^{+3.0} _{-2.2} $ & $0.95^{+1.13} _{-0.52}$ &	$-12.02 \pm 0.34$ \\
SDSS J2225-0014	& 4.886	      & 0.005      & \Lya forest & $2.144	\pm 0.885$   & --27.39	  &	$6.0 ^{+4.2} _{-2.5} $ & $1.02^{+2.51} _{-0.72}$ &	$-11.99 \pm 0.54$ \\
SDSS J1616+0501	& 4.876	      & 0.005      & \Lya forest & $2.603	\pm 1.020$   & --27.60	  &	$3.3 ^{+2.6} _{-1.5} $ & $3.89^{+10.80}_{-2.86}$ &	$-11.41 \pm 0.58$ \\
SDSS J2147-0838	& 4.588	      & 0.005      & \Lya forest & $3.506	\pm 1.245$   & --27.93	  &	$5.3 ^{+4.0} _{-2.3} $ & $1.88^{+4.58} _{-1.33}$ &	$-11.73 \pm 0.54$ \\

\hline
\end{tabular}
\label{tab:results}
\end{center}
\end{minipage}
\end{table*}

\subsection{Results}   
\label{sect:results}

Table~\ref{tab:results} summarises the main
results of this paper for the individual
quasar sightlines. The quoted errors in
$\Gamma_{\rm bkg}$ consist of the measured
error from $10\,000$ Monte Carlo realisations
of the spectra, varying the luminosity,
systemic redshift and forest $\tau_{\rm eff}$,
added in quadrature to the expected
sightline-to-sightline scatter as
described in Section~\ref{sect:reliable}.
The spectra with the largest errors in
$L^{\rm Q}_{\nu_0}$ are those with continuum
magnitudes measured directly from the SDSS
spectra. Even though the errors in the
luminosity could be rather large ($\sim 30$
per cent), the measured error was strongly
dominated by errors in the redshift and
$\tau_{\rm eff}$. The statistical power of a
measurement from a single spectrum is rather
limited and the error on $\Gamma_{\rm bkg}$ 
is substantial. Fig.~\ref{fig:results} shows
the estimated $\log(\Gamma_{\rm bkg})$ and
$R_{\rm eq}$ for our sample as a function of
redshift. The data points in Fig.~\ref{fig:results}
are consistent with a linear fit within the errors.
There is therefore little we can say about
spatial fluctuations in UVB intensity other
than that they appear to be smaller than our
measurement errors. We note that if we had
used the $\alpha$ presented in \citet{Scott2004a}
then the $\log(\Gamma_{\rm bkg})$ values would
have been on average 0.22 dex higher, and the
error bars 10 per cent larger. We also note that
our redshift evolution of $R_{\rm eq}$ is opposite
to the redshift evolution of the proximity region
sizes as presented by \citet{Fan2006c}. Note that
this is entirely due to the different definition
of the two sizes considered. For the proximity
region size definition from \citet{Fan2006c} we
find similar results.

\begin{figure*}
	\centering
		\includegraphics[width=17cm]{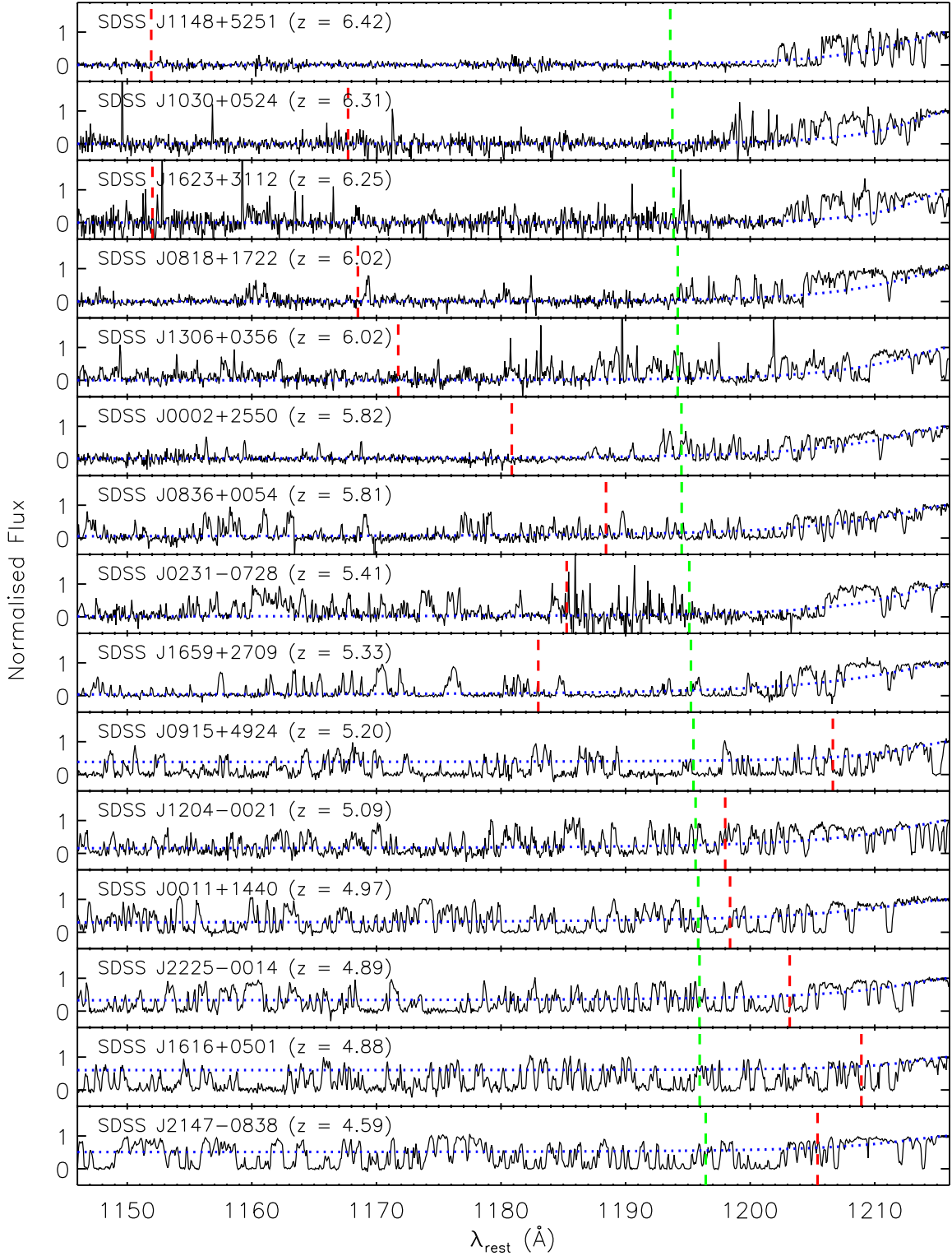}
	\caption{
	The 15 spectra used in this study, ordered
	by redshift. The spectra are presented in
	their rest wavelength, $\lambda_{\rm rest}$,
	in order to emphasise the proximity effect
	region bluewards of the \Lya emission line,
  on the far right of the plot at 1216~\AA.
  They have been normalised with the continuum
  fitting process described in
  Section~\ref{sect:obs_spec}, and have also
  been smoothed to a common pixel size of
  $16.7\kmps$ (observed frame) for clarity.
  The red dashed line indicates the derived
  value of $R_{\rm eq}$, whilst the blue
  dotted line is the expected fall-off in
  average flux. The area of spectrum to the
  right of the green dashed line was the
  section used for the proximity effect
  measurement, and is $40h^{-1}$ comoving
  Mpc long.
	}
	\label{fig:hires-mike}
\end{figure*}

In Fig.~\ref{fig:lum_ind} we subtract out 
the linear evolution of the average  
$\log(\Gamma_{\rm bkg})$. There is no
systematic trend with the luminosity of the
quasar. This confirms that the systematic
shifts induced by the luminosity of the
quasar (see Table~\ref{tab:err_budget} and
Fig.~\ref{fig:hires_req}) have been suitably
corrected for. Fig.~\ref{fig:hires-mike}
shows a sample of the observed spectra with
their derived $R_{\rm eq}$ and expected
average flux fall-off, in their rest frame,
as well as denoting the section of each
spectrum used for the measurement.
Judging by visual inspection, none of our
observed spectra appears to be influenced
by LLSs to the degree presented in the
bottom panel of Fig.~\ref{fig:shortening}.
It is, however, difficult to determine if
any of the sightlines contained LLSs far
from the quasar redshift, as in the middle
panel of Fig.~\ref{fig:shortening}.
SDSS J1148+5251 has a very short region of
enhanced transmitted flux (Fig.~\ref{fig:hires-mike})
and is a possible candidate for being
affected by a LLS.

Our most robust results come from
simultaneously fitting multiple sightlines.
The data were grouped into low redshift
($z < 5.5$) and high redshift ($z > 5.5$)
samples, and $\Gamma_{\rm bkg}$ was measured
by finding where $\sum_i{\Delta F_i}=0$,
with $\Delta F$ calculated for each spectrum
using the same trial $\Gamma_{\rm bkg}$.
For the lower redshift bin, containing 8
spectra, the average redshift was $z=5.04$
and gave
$\log(\Gamma_{\rm bkg}) = -12.15 \pm 0.16~(0.32)$
at 68 (95) per cent confidence, whilst for
the higher redshift bin with 7 spectra and
an average redshift of $z=6.09$,
$\log(\Gamma_{\rm bkg}) = -12.84 \pm 0.18~(0.36)$.
These will subsequently be referred to as
the $z \sim 5$ and $z \sim 6$ samples,
respectively, for ease of comparison to the
simulations and previous work. The implications
of these results will now be discussed.

\subsection{Comparison to previous work}
\label{sect:comp}
\begin{figure*}
	\centering
   \includegraphics[width=18cm]{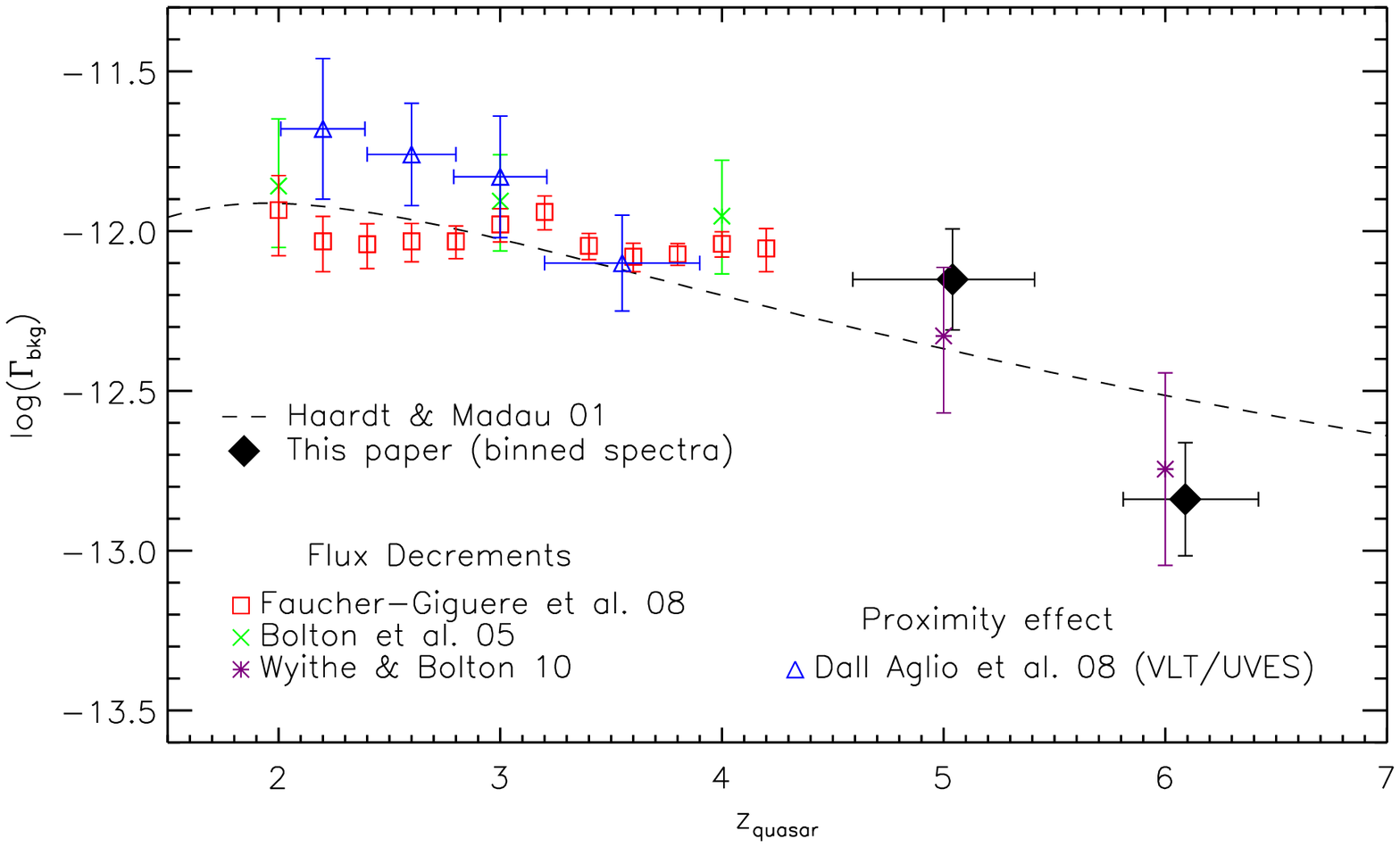}
	\caption{
	The evolution of the UVB in the redshift
	range $z=2$ to $z=6$. The solid points are
	the results of this paper, and represent the
	aggregate analysis over multiple sightlines.
	The spectra were coarsely binned into two
	subsets, namely $z > 5.5$ and $z < 5.5$. The
	lower redshift subset contained 8 spectra and
	had an average redshift of $z_{\rm av} = 5.04$
	whilst the upper had 7 spectra and
	$z_{\rm av} = 6.09$. The errors displayed on
	our points are	the 68 per cent confidence
	intervals, and exclude any correction due
	to the presence of LLS,	local overdensities,
	or a thermal proximity effect. Recent
	estimates from the literature are also plotted
	with their $1 \sigma$ errors.	The flux
	decrement results have all been scaled to our
	adopted cosmology
	$(h, \Omega_m, \Omega_b h^2, \sigma_8) = (0.72,0.26,0.024,0.85)$
	and to the same temperature-density relation,
	$T = T_0 \Delta^{\gamma-1}$, with $T_0$ and
	$\gamma$ held constant at $10^4$~K and 1.3
	respectively, using	the scaling relations
	from \citet{Bolton2005a} and \citet{Bolton2007a}.
	Our results at both $z \sim 5$ and $z \sim 6$
	agree very well with those from flux decrement
	measurements.	The theoretical curve of
	\citet{Haardt2001a}, assuming	contributions
	from quasars and star forming	galaxies,	is also
	plotted. 
	}
	\label{fig:z-gamma-others-binned}
\end{figure*}

The results of fitting multiple lines of
sight with a constant UVB are plotted in
Fig.~\ref{fig:z-gamma-others-binned}.
We also plot a selection of UVB estimates
from the literature, where the literature
results have been scaled to the cosmology
used in this paper, as well as the same
temperature-density relation parameters
used by \citet{Bolton2007a}, such that
$T = T_0 \Delta^{\gamma-1}$, with $T_0$
and $\gamma$	held constant at $10^4$~K
and 1.3 respectively. The UVB model of
\citet{Haardt2001a} is also shown. As
previously noted in Fig.~\ref{fig:results},
whilst the error bars are large for the
individual sightlines, there is a clear
trend of a decreasing UVB intensity in
the redshift range $z \sim 5-6$. This
decrease is more pronounced in the
binned results.

The results presented here are the first
proximity effect measurements of
$\Gamma_{\rm bkg}$ at these redshifts.
\citet{DallAglio2009b} detected the
proximity effect in 1733 spectra from the
SDSS over the range $2 \lesssim z \lesssim 4.5$
and found that the UVB seemed to be
remarkably flat over this redshift range, with
$\log(\Gamma_{\rm bkg}) = -11.78 \pm 0.07$.
This value is consistent with our results
from individual sightlines over $4.5 < z < 5$.
A more direct comparison can be made with
the results of \citet{DallAglio2008a}, who
measured the proximity effect in 40 spectra
from the UVES instrument on the VLT.
Comparing our results to the \citet{DallAglio2008a}
results may be more appropriate, as both
studies used high-resolution spectra. Combined,
the proximity effect results suggest a smooth
decrease in UVB from $z \sim 2$ to $z \sim 6$
by an order of magnitude
(Fig.~\ref{fig:z-gamma-others-binned}).

The range $z \sim 5-6$ has previously
only been probed using flux decrements
\citep{McDonald2001a, Meiksin2004a, Bolton2007a, Wyithe2010a}.
There is excellent agreement between our
binned results and those from flux decrements
(see Fig.~\ref{fig:z-gamma-others-binned}),
which also suggest a significant decline
in the UVB from $z \sim 4$ to $z \sim 6$.

Our measured evolution of the UVB intensity has
important implications for reionization. The \HI
photoionisation rate should scale as
$\Gamma(z) \propto l(\nu_0,z)\epsilon_{\nu_0}$,
where $l(\nu_0,z)$ is the mean free path of
ionising photons and $\epsilon_{\nu_0}$ is the
ionising emissivity. The redshift evolution of
$\Gamma_{\rm bkg}$ therefore gives an insight
into the evolution of these two key variables.
During the `percolation' stage at the end of
reionization, $l(\nu_0,z)$ is expected to evolve
rapidly, in marked contrast to its gradual
evolution in the  post-reionization Universe
\citep{Gnedin2006a}. Consequently, the end of
reionization should be indicated by a break in
the evolution of $l(\nu_0,z)$. The smooth
redshift evolution of our measurements of the 
UVB intensity (see left panel of Fig.~\ref{fig:results})
implies that both $l(\nu_0,z)$ and
$\epsilon_{\nu_0}$ are also evolving smoothly
in the redshift range $4.6<z<6.4$, as otherwise
they would have to both evolve rapidly
simultaneously in opposite directions, which
appears very unlikely. This suggests that
percolation has occurred at higher redshifts
than are probed by our sample
\citep[although see][]{Furlanetto2009a}.

Our measurements of the UVB can be combined
with measurements of the mean free path to
place constraints on the evolution of the
ionising emissivity. \citet{Songaila2010a}
recently measured the incidence of LLS over
$0<z<6$. They find an evolution in the mean
free path which can be approximated as 
$l(\nu_0,z) = 50[(1+z)/4.5]^{-4.44^{+0.36}_{-0.32}}$.
Using this fit, we infer that $l(\nu_0,z)$
decreases by a factor $\sim 1.5-2.5$ from
$z = 5-6$. At face value the decrease of our
measurements of $\Gamma_{\rm bkg}$ with
redshift (a factor of $\sim 2.5-8$) therefore
imply an emissivity that is either roughly
constant or drops by up to a factor of $\sim 5$.
Note that this is consistent with the decrease
in the (dust-corrected) UV luminosity density
between $z=5-6$ measured by \citet{Bouwens2009a}
for rest wavelength $\sim 1600$~\AA~and integrated
to either $0.3 L^*_{z=3}$ or $0.04 L^*_{z=3}$
(where $L^*_{z=3}$ is the luminosity derived
by \citet{Steidel1999a} at $z \sim 3$ and
corresponds to $M_{1700,AB} = -21.07$).

The fit to the mean free path from
\citet{Songaila2010a} is plotted with
the measured proximity region sizes in the
right panel of Fig.~\ref{fig:results}.
Our proximity region sizes become larger
than the mean free path as measured by
\citet{Songaila2010a} at $z \gtrsim 5.5$.
If the quasar is able to ionise LLS
within the proximity region, then the
mean free path within the proximity
region will increase, allowing an enhanced
contribution to the ionisation rate from
local galaxies and AGN. As a result, the
proximity region may appear larger, and so
the UVB may be underestimated in a proximity
effect analysis. Careful modeling of LLS
will be required to determine whether this
is an important effect. However, it may partly
explain the decrease in ionising emissivity
we infer from $z \sim 5$ to 6
(see also Section~\ref{sect:halo-mass}).

The ionising emissivity is already very low
at $z=6$ ($\sim 1.5$ ionising photons per
hydrogen atom) and so this evolution can not
continue to much higher redshifts without
reionization failing to complete by $z=6$
\citep{Bolton2007a}. For this reason, unless
there is a substantial increase in
$\epsilon_{\nu_0}$ at very high redshifts,
the end of reionization, while potentially
before $z=6.4$, appears unlikely to occur
much earlier.

\section{Conclusions}
\label{sect:conc}
We have presented new measurements of the
photoionisation rate of hydrogen by the
ultraviolet background (UVB) using the
proximity effect in quasar spectra. The
fifteen spectra in the sample cover the range
$4.6<z<6.4$, allowing us to conduct the
first proximity effect measurements of
$\Gamma_{\rm bkg}$ at these high redshifts.

For each quasar, $\Gamma_{\rm bkg}$ was
calculated by modelling the total
photoionisation rate as a function of
distance from the quasar, taking into
account contributions from both the quasar
and the UVB. The optical depths in the
proximity zone were then modified to
produce the expected optical depths in
the absence of the enhanced ionisation
from the quasar. The preferred value of
$\Gamma_{\rm bkg}$ was the one for which
the resulting mean flux in the proximity
zone was equal to that of the average
\Lya forest at the same redshift.

We investigated a wide range of potential
errors and biases affecting the proximity
effect measurements using full radiative
transfer simulations. We found the error
in $\Gamma_{\rm bkg}$ for an individual
line of sight to be dominated by the error
in the quasar redshift and $\tau_{\rm eff}$.
The redshift errors for the sample are
generally $\leq 0.01$, with several having
a very accurate systemic redshift from CO
and \MgII. An error in the assumed
luminosity of the quasar of up to 40 per
cent would still have a smaller effect 
than the typical redshift errors. Our raw
measurements of the UVB intensity from
individual sight lines at $z=5$ should be 
typically overestimated by a factor of two,
with random errors that are also roughly a
factor of two. The largest biases result
from the finite $S/N$ of the spectra and
the quasar lying in an overdensity
(50 and 30 per cent, respectively).
At $z=6$, the UVB is overestimated by 30
per cent, with a factor of 2.5 random
error. The largest source of bias at this
redshift is the uncertainty in
$\tau_{\rm eff}$.

Systematic uncertainties caused by the
overdensity of matter close to the quasar
were found to be smaller at $z>5$ than at
lower redshifts. The effect of the quasars
lying in large-scale overdensities was
found to be small for host halo masses of
$M \leq 10^{12.2} \Msun$, the most massive
halos in our largest ($80h^{-1}$ comoving
Mpc) simulation. Even in a toy model
emulating the density distribution
surrounding a $10^{13} \Msun$ halo at
$z = 6$, (and therefore as massive as the
most massive haloes in the Millennium
simulation at this redshift) large-scale
overdensities were not dominating the error
budget. This is due to the proximity
regions (as defined in this paper) being
substantially larger than the scale of
the overdensities. It is however noted
that at lower redshifts, where the UVB
intensity and the mean flux level are
higher, proximity regions may well be of
similar size to the local overdensity.
This effect is therefore likely very
important at $z \sim 2-3$, where
discrepancies between measurements from
flux decrements and the proximity effect
have been noted
\citep{Rollinde2005a,Guimaraes2007a}.

We also tested the effect of Lyman limit
systems (LLS) Using the recent measurements
of the frequency of LLS at high redshift from
\citet{Songaila2010a}, we found that LLS have
also only a small effect on our analysis,
potentially causing us to overestimate
$\Gamma_{\rm bkg}$ in our combined samples at
$z \sim 5$ and 6 by roughly 0.1 dex.

Our measurements further assume that the
gas temperature within the proximity region
is similar to that in the general IGM. A
thermal proximity effect, could have a
sizeable impact on our estimates of the 
UVB intensity. If the gas in the $\sim 5$
proper Mpc of the proximity region closest
to the quasar were e.g. on average $10^4$~K
hotter than the general IGM then the UVB will
be {\it{underestimated}} by $0.14~(0.21)$ dex
at $z=5~(6)$. This is comparable in magnitude
to the biases from other environmental
factors, although opposite in sign.

The results quoted in this paper are corrected
for these biases, except the environmental
biases due to a different gas temperature and
average gas density, and the presence of LLS
in the proximity zone. We have, however,
demonstrated these biases are small or, in
case of the temperature, are poorly
observationally constrained.

Our measured values of $\Gamma_{\rm bkg}$,
corrected for biases, decline significantly
from $z \sim 5$ to 6. For $z \sim 5$ we find
$\log(\Gamma_{\rm bkg}) = -12.15 \pm 0.16~(0.32)$
at 68 (95) per cent confidence, whilst at $z \sim 6$
we find
$\log(\Gamma_{\rm bkg}) = -12.77 \pm 0.18~(0.36)$,
a decline significant at roughly the $3 \sigma$
level. Within our sample, the UVB intensity
measured from individual sight lines is seen to
decline smoothly with redshift over $4.6 < z < 6.4$,
but show no sign of the rapid decline which may
be expected in the late stages of reionization 
when there is a rapid change in the attenuation
length of ionising photons.

Our results are in good agreement with UVB
estimates from measurements of the mean flux
decrement in the redshift range $z=5-6$,
assuming an IGM temperature $T_0 = 10^4$~K.
Both the proximity effect and flux decrement
measurements imply a decline in the intensity
of the UVB by nearly an order of magnitude
from $z=4$ to $z=6$.

Finally, we have combined our estimates of
$\Gamma_{\rm bkg}$ with the evolution of the
mean free path measured by \citet{Songaila2010a}.
At face value the combined measurements imply
a decline in the ionising emissivity of a factor
of about 2 from $z \sim 5$ to 6, but it is
important to keep in mind that at these redshifts
measurements of the mean free path of ionising
photons are extremely difficult, and that with
such a low emissivity reionization could barely
have been completed by $z = 6.4$. 

The results presented here represent some of
the highest redshift measurements of the UV
background made to date, enabling us to probe
deeper into the late(st) stages of hydrogen
reionization. The next generation of optical
and NIR telescopes will enable access to high
signal-to-noise quasar spectra at even higher
redshifts, leading to improved measurements
of the ionising background and helping to
establish a more complete picture of the final
stages of the hydrogen reionization epoch.

\section*{Acknowledgments}
The authors would like to thank the anonymous
referee for helpful comments, and Max Pettini
and Richard McMahon for constructive and useful
conversations during the course of this study.
The authors are grateful to the staff at Keck
and Magellan for making these observations
possible. We also wish to recognise and
acknowledge the very significant cultural role
and reverence that the summit of Mauna Kea has
always had within the indigenous Hawaiian
community. We are most fortunate to have the
opportunity to conduct observations from this
mountain. The hydrodynamical simulations used
in this work were performed in cooperation
with SGI/Intel using the COSMOS facility
hosted by the Department of Applied
Mathematics and Theoretical Physics (DAMTP)
at the University of Cambridge. COSMOS is a
UK-CCC facility supported The Higher Education
Funding Council for England (HEFCE) and the
UK Science and Technology Facilities Council
(STFC). We would like to thank STFC for
financial support. GB acknowledges financial
support from the Kavli foundation. JB has been
supported by an ARC Australian postdoctoral
fellowship (DP0984947).

\bibliographystyle{mnras}
\bibliography{cbhb10_bib}

\begin{thebibliography}{}

\bibitem[\protect\citeauthoryear{{Adelberger} et~al.}{{Adelberger}
  et~al.}{2003}]{Adelberger2003a}
{Adelberger} K.~L., {Steidel} C.~C., {Shapley} A.~E.,  {Pettini} M., 2003,
  \apj, 584, 45

\bibitem[\protect\citeauthoryear{{Bajtlik}, {Duncan}, \& {Ostriker}}{{Bajtlik}
  et~al.}{1988}]{Bajtlik1988a}
{Bajtlik} S., {Duncan} R.~C.,  {Ostriker} J.~P., 1988, \apj, 327, 570

\bibitem[\protect\citeauthoryear{{Bechtold} et~al.}{{Bechtold}
  et~al.}{1987}]{Bechtold1987a}
{Bechtold} J., {Weymann} R.~J., {Lin} Z.,  {Malkan} M.~A., 1987, \apj, 315, 180

\bibitem[\protect\citeauthoryear{{Becker} et~al.}{{Becker}
  et~al.}{2010}]{Becker2010a}
{Becker} G.~D., {Bolton} J.~S., {Haehnelt} M.~G.,  {Sargent} W.~L.~W., 2010,
  ArXiv e-prints

\bibitem[\protect\citeauthoryear{{Becker}, {Rauch}, \& {Sargent}}{{Becker}
  et~al.}{2007}]{Becker2007a}
{Becker} G.~D., {Rauch} M.,  {Sargent} W.~L.~W., 2007, \apj, 662, 72

\bibitem[\protect\citeauthoryear{{Becker} et~al.}{{Becker}
  et~al.}{2006}]{Becker2006a}
{Becker} G.~D., {Sargent} W.~L.~W., {Rauch} M.,  {Simcoe} R.~A., 2006, \apj,
  640, 69

\bibitem[\protect\citeauthoryear{{Bernstein} et~al.}{{Bernstein}
  et~al.}{2003}]{Bernstein2003a}
{Bernstein} R., {Shectman} S.~A., {Gunnels} S.~M., {Mochnacki} S.,  {Athey}
  A.~E., 2003, in Presented at the Society of Photo-Optical Instrumentation
  Engineers (SPIE) Conference, Vol. 4841, {M.~Iye \& A.~F.~M.~Moorwood} , ed,
  Society of Photo-Optical Instrumentation Engineers (SPIE) Conference Series,
  p. 1694

\bibitem[\protect\citeauthoryear{{Bolton} \& {Becker}}{{Bolton} \&
  {Becker}}{2009}]{Bolton2009a}
{Bolton} J.~S.,  {Becker} G.~D., 2009, \mnras, 398, L26

\bibitem[\protect\citeauthoryear{{Bolton} et~al.}{{Bolton}
  et~al.}{2010}]{Bolton2010a}
{Bolton} J.~S., {Becker} G.~D., {Wyithe} J.~S.~B., {Haehnelt} M.~G.,  {Sargent}
  W.~L.~W., 2010, \mnras, 406, 612

\bibitem[\protect\citeauthoryear{{Bolton} \& {Haehnelt}}{{Bolton} \&
  {Haehnelt}}{2007a}]{Bolton2007b}
{Bolton} J.~S.,  {Haehnelt} M.~G., 2007a, \mnras, 374, 493

\bibitem[\protect\citeauthoryear{{Bolton} \& {Haehnelt}}{{Bolton} \&
  {Haehnelt}}{2007b}]{Bolton2007a}
{Bolton} J.~S.,  {Haehnelt} M.~G., 2007b, \mnras, 382, 325

\bibitem[\protect\citeauthoryear{{Bolton} et~al.}{{Bolton}
  et~al.}{2005}]{Bolton2005a}
{Bolton} J.~S., {Haehnelt} M.~G., {Viel} M.,  {Springel} V., 2005, \mnras, 357,
  1178

\bibitem[\protect\citeauthoryear{{Bonoli} et~al.}{{Bonoli}
  et~al.}{2009}]{Bonoli2009a}
{Bonoli} S., {Marulli} F., {Springel} V., {White} S.~D.~M., {Branchini} E.,
  {Moscardini} L., 2009, \mnras, 396, 423

\bibitem[\protect\citeauthoryear{{Bouwens} et~al.}{{Bouwens}
  et~al.}{2006}]{Bouwens2006a}
{Bouwens} R.~J., {Illingworth} G.~D., {Blakeslee} J.~P.,  {Franx} M., 2006,
  \apj, 653, 53

\bibitem[\protect\citeauthoryear{{Bouwens} et~al.}{{Bouwens}
  et~al.}{2009}]{Bouwens2009a}
{Bouwens} R.~J. et~al., 2009, \apj, 705, 936

\bibitem[\protect\citeauthoryear{{Bouwens} et~al.}{{Bouwens}
  et~al.}{2008}]{Bouwens2008a}
{Bouwens} R.~J., {Illingworth} G.~D., {Franx} M.,  {Ford} H., 2008, \apj, 686,
  230

\bibitem[\protect\citeauthoryear{{Bouwens} et~al.}{{Bouwens}
  et~al.}{2010}]{Bouwens2010a}
{Bouwens} R.~J. et~al., 2010, ArXiv e-prints

\bibitem[\protect\citeauthoryear{{Bunker} et~al.}{{Bunker}
  et~al.}{2004}]{Bunker2004a}
{Bunker} A.~J., {Stanway} E.~R., {Ellis} R.~S.,  {McMahon} R.~G., 2004, \mnras,
  355, 374

\bibitem[\protect\citeauthoryear{{Carilli} et~al.}{{Carilli}
  et~al.}{2010}]{Carilli2010a}
{Carilli} C.~L. et~al., 2010, \apj, 714, 834

\bibitem[\protect\citeauthoryear{{Carswell} et~al.}{{Carswell}
  et~al.}{1987}]{Carswell1987a}
{Carswell} R.~F., {Webb} J.~K., {Baldwin} J.~A.,  {Atwood} B., 1987, \apj, 319,
  709

\bibitem[\protect\citeauthoryear{{Carswell} et~al.}{{Carswell}
  et~al.}{1982}]{Carswell1982a}
{Carswell} R.~F., {Whelan} J.~A.~J., {Smith} M.~G., {Boksenberg} A.,  {Tytler}
  D., 1982, \mnras, 198, 91

\bibitem[\protect\citeauthoryear{{Cooke}, {Espey}, \& {Carswell}}{{Cooke}
  et~al.}{1997}]{Cooke1997a}
{Cooke} A.~J., {Espey} B.,  {Carswell} R.~F., 1997, \mnras, 284, 552

\bibitem[\protect\citeauthoryear{{Cristiani} et~al.}{{Cristiani}
  et~al.}{1995}]{Cristiani1995a}
{Cristiani} S., {D'Odorico} S., {Fontana} A., {Giallongo} E.,  {Savaglio} S.,
  1995, \mnras, 273, 1016

\bibitem[\protect\citeauthoryear{{da {\^A}ngela} et~al.}{{da {\^A}ngela}
  et~al.}{2008}]{daAngela2008a}
{da {\^A}ngela} J. et~al., 2008, \mnras, 383, 565

\bibitem[\protect\citeauthoryear{{Dall'Aglio} \& {Gnedin}}{{Dall'Aglio} \&
  {Gnedin}}{2010}]{DallAglio2010a}
{Dall'Aglio} A.,  {Gnedin} N.~Y., 2010, \apj, 722, 699

\bibitem[\protect\citeauthoryear{{Dall'Aglio}, {Wisotzki}, \&
  {Worseck}}{{Dall'Aglio} et~al.}{2008}]{DallAglio2008a}
{Dall'Aglio} A., {Wisotzki} L.,  {Worseck} G., 2008, \aap, 491, 465

\bibitem[\protect\citeauthoryear{{Dall'Aglio}, {Wisotzki}, \&
  {Worseck}}{{Dall'Aglio} et~al.}{2009}]{DallAglio2009b}
{Dall'Aglio} A., {Wisotzki} L.,  {Worseck} G., 2009, ArXiv e-prints

\bibitem[\protect\citeauthoryear{{D'Odorico} et~al.}{{D'Odorico}
  et~al.}{2008}]{DOdorico2008a}
{D'Odorico} V., {Bruscoli} M., {Saitta} F., {Fontanot} F., {Viel} M.,
  {Cristiani} S.,  {Monaco} P., 2008, \mnras, 389, 1727

\bibitem[\protect\citeauthoryear{{Eisenstein} \& {Hu}}{{Eisenstein} \&
  {Hu}}{1999}]{Eisenstein1999a}
{Eisenstein} D.~J.,  {Hu} W., 1999, \apj, 511, 5

\bibitem[\protect\citeauthoryear{{Espey}}{{Espey}}{1993}]{Espey1993a}
{Espey} B.~R., 1993, \apjl, 411, L59

\bibitem[\protect\citeauthoryear{{Fan} et~al.}{{Fan} et~al.}{2004}]{Fan2004a}
{Fan} X. et~al., 2004, \aj, 128, 515

\bibitem[\protect\citeauthoryear{{Fan} et~al.}{{Fan} et~al.}{2001}]{Fan2001a}
{Fan} X. et~al., 2001, \aj, 122, 2833

\bibitem[\protect\citeauthoryear{{Fan} et~al.}{{Fan} et~al.}{2006a}]{Fan2006c}
{Fan} X. et~al., 2006a, \aj, 132, 117

\bibitem[\protect\citeauthoryear{{Fan} et~al.}{{Fan} et~al.}{2006b}]{Fan2006a}
{Fan} X. et~al., 2006b, \aj, 131, 1203

\bibitem[\protect\citeauthoryear{{Fan} et~al.}{{Fan} et~al.}{2003}]{Fan2003a}
{Fan} X. et~al., 2003, \aj, 125, 1649

\bibitem[\protect\citeauthoryear{{Fardal}, {Giroux}, \& {Shull}}{{Fardal}
  et~al.}{1998}]{Fardal1998a}
{Fardal} M.~A., {Giroux} M.~L.,  {Shull} J.~M., 1998, \aj, 115, 2206

\bibitem[\protect\citeauthoryear{{Faucher-Gigu{\`e}re}
  et~al.}{{Faucher-Gigu{\`e}re} et~al.}{2008}]{Faucher2008a}
{Faucher-Gigu{\`e}re} C., {Lidz} A., {Zaldarriaga} M.,  {Hernquist} L., 2008,
  \apj, 673, 39

\bibitem[\protect\citeauthoryear{{Fontanot} et~al.}{{Fontanot}
  et~al.}{2006}]{Fontanot2006a}
{Fontanot} F., {Monaco} P., {Cristiani} S.,  {Tozzi} P., 2006, \mnras, 373,
  1173

\bibitem[\protect\citeauthoryear{{Furlanetto} \& {Mesinger}}{{Furlanetto} \&
  {Mesinger}}{2009}]{Furlanetto2009a}
{Furlanetto} S.~R.,  {Mesinger} A., 2009, \mnras, 394, 1667

\bibitem[\protect\citeauthoryear{{Gnedin} \& {Fan}}{{Gnedin} \&
  {Fan}}{2006}]{Gnedin2006a}
{Gnedin} N.~Y.,  {Fan} X., 2006, \apj, 648, 1

\bibitem[\protect\citeauthoryear{{Granato} et~al.}{{Granato}
  et~al.}{2004}]{Granato2004a}
{Granato} G.~L., {De Zotti} G., {Silva} L., {Bressan} A.,  {Danese} L., 2004,
  \apj, 600, 580

\bibitem[\protect\citeauthoryear{{Guimar{\~a}es} et~al.}{{Guimar{\~a}es}
  et~al.}{2007}]{Guimaraes2007a}
{Guimar{\~a}es} R., {Petitjean} P., {Rollinde} E., {de Carvalho} R.~R.,
  {Djorgovski} S.~G., {Srianand} R., {Aghaee} A.,  {Castro} S., 2007, \mnras,
  377, 657

\bibitem[\protect\citeauthoryear{{Haardt} \& {Madau}}{{Haardt} \&
  {Madau}}{1996}]{Haardt1996a}
{Haardt} F.,  {Madau} P., 1996, \apj, 461, 20

\bibitem[\protect\citeauthoryear{{Haardt} \& {Madau}}{{Haardt} \&
  {Madau}}{2001}]{Haardt2001a}
{Haardt} F.,  {Madau} P., 2001, in {D.~M.~Neumann \& J.~T.~V.~Tran} , ed,
  Clusters of Galaxies and the High Redshift Universe Observed in X-rays

\bibitem[\protect\citeauthoryear{{Jena} et~al.}{{Jena}
  et~al.}{2005}]{Jena2005a}
{Jena} T. et~al., 2005, \mnras, 361, 70

\bibitem[\protect\citeauthoryear{{Kim} \& {Croft}}{{Kim} \&
  {Croft}}{2007}]{Kim2007a}
{Kim} Y.,  {Croft} R., 2007, ArXiv Astrophysics e-prints

\bibitem[\protect\citeauthoryear{{Kirkman} \& {Tytler}}{{Kirkman} \&
  {Tytler}}{2008}]{Kirkman2008a}
{Kirkman} D.,  {Tytler} D., 2008, \mnras, 391, 1457

\bibitem[\protect\citeauthoryear{{Kirkman} et~al.}{{Kirkman}
  et~al.}{2005}]{Kirkman2005a}
{Kirkman} D. et~al., 2005, \mnras, 360, 1373

\bibitem[\protect\citeauthoryear{{Komatsu} et~al.}{{Komatsu}
  et~al.}{2009}]{Komatsu2009a}
{Komatsu} E. et~al., 2009, \apjs, 180, 330

\bibitem[\protect\citeauthoryear{{Kulkarni} \& {Fall}}{{Kulkarni} \&
  {Fall}}{1993}]{Kulkarni1993a}
{Kulkarni} V.~P.,  {Fall} S.~M., 1993, \apjl, 413, L63

\bibitem[\protect\citeauthoryear{{Lidz} et~al.}{{Lidz}
  et~al.}{2007}]{Lidz2007a}
{Lidz} A., {McQuinn} M., {Zaldarriaga} M., {Hernquist} L.,  {Dutta} S., 2007,
  \apj, 670, 39

\bibitem[\protect\citeauthoryear{{Liske} \& {Williger}}{{Liske} \&
  {Williger}}{2001}]{Liske2001a}
{Liske} J.,  {Williger} G.~M., 2001, \mnras, 328, 653

\bibitem[\protect\citeauthoryear{{Loeb} \& {Eisenstein}}{{Loeb} \&
  {Eisenstein}}{1995}]{Loeb1995a}
{Loeb} A.,  {Eisenstein} D.~J., 1995, \apj, 448, 17

\bibitem[\protect\citeauthoryear{{Lu} et~al.}{{Lu} et~al.}{1996}]{Lu1996a}
{Lu} L., {Sargent} W.~L.~W., {Womble} D.~S.,  {Takada-Hidai} M., 1996, \apj,
  472, 509

\bibitem[\protect\citeauthoryear{{Maselli}, {Ferrara}, \&
  {Gallerani}}{{Maselli} et~al.}{2009}]{Maselli2009a}
{Maselli} A., {Ferrara} A.,  {Gallerani} S., 2009, \mnras, 395, 1925

\bibitem[\protect\citeauthoryear{{McDonald} \& {Miralda-Escud{\'e}}}{{McDonald}
  \& {Miralda-Escud{\'e}}}{2001}]{McDonald2001a}
{McDonald} P.,  {Miralda-Escud{\'e}} J., 2001, \apjl, 549, L11

\bibitem[\protect\citeauthoryear{{Meiksin} \& {White}}{{Meiksin} \&
  {White}}{2004}]{Meiksin2004a}
{Meiksin} A.,  {White} M., 2004, \mnras, 350, 1107

\bibitem[\protect\citeauthoryear{{Mesinger} \& {Furlanetto}}{{Mesinger} \&
  {Furlanetto}}{2009}]{Mesinger2009a}
{Mesinger} A.,  {Furlanetto} S., 2009, \mnras, 400, 1461

\bibitem[\protect\citeauthoryear{{Miralda-Escud{\'e}}}{{Miralda-Escud{\'e}}}{2%
003}]{Miralda2003a}
{Miralda-Escud{\'e}} J., 2003, \apj, 597, 66

\bibitem[\protect\citeauthoryear{{Miralda-Escud{\'e}}, {Haehnelt}, \&
  {Rees}}{{Miralda-Escud{\'e}} et~al.}{2000}]{Miralda2000a}
{Miralda-Escud{\'e}} J., {Haehnelt} M.,  {Rees} M.~J., 2000, \apj, 530, 1

\bibitem[\protect\citeauthoryear{{Murdoch} et~al.}{{Murdoch}
  et~al.}{1986}]{Murdoch1986a}
{Murdoch} H.~S., {Hunstead} R.~W., {Pettini} M.,  {Blades} J.~C., 1986, \apj,
  309, 19

\bibitem[\protect\citeauthoryear{{Oesch} et~al.}{{Oesch}
  et~al.}{2010}]{Oesch2010a}
{Oesch} P.~A. et~al., 2010, \apjl, 709, L16

\bibitem[\protect\citeauthoryear{{Ouchi} et~al.}{{Ouchi}
  et~al.}{2009}]{Ouchi2009a}
{Ouchi} M. et~al., 2009, \apj, 706, 1136

\bibitem[\protect\citeauthoryear{{Pascarelle} et~al.}{{Pascarelle}
  et~al.}{2001}]{Pascarelle2001a}
{Pascarelle} S.~M., {Lanzetta} K.~M., {Chen} H.,  {Webb} J.~K., 2001, \apj,
  560, 101

\bibitem[\protect\citeauthoryear{{Pentericci} et~al.}{{Pentericci}
  et~al.}{2002}]{Pentericci2002a}
{Pentericci} L. et~al., 2002, \aj, 123, 2151

\bibitem[\protect\citeauthoryear{{P{\'e}roux} et~al.}{{P{\'e}roux}
  et~al.}{2003}]{Peroux2003a}
{P{\'e}roux} C., {McMahon} R.~G., {Storrie-Lombardi} L.~J.,  {Irwin} M.~J.,
  2003, \mnras, 346, 1103

\bibitem[\protect\citeauthoryear{{Prochaska}, {O'Meara}, \&
  {Worseck}}{{Prochaska} et~al.}{2010}]{Prochaska2010a}
{Prochaska} J.~X., {O'Meara} J.~M.,  {Worseck} G., 2010, \apj, 718, 392

\bibitem[\protect\citeauthoryear{{Rauch} et~al.}{{Rauch}
  et~al.}{1997}]{Rauch1997a}
{Rauch} M. et~al., 1997, \apj, 489, 7

\bibitem[\protect\citeauthoryear{{Richard} et~al.}{{Richard}
  et~al.}{2006}]{Richard2006a}
{Richard} J., {Pell{\'o}} R., {Schaerer} D., {Le Borgne} J.,  {Kneib} J., 2006,
  \aap, 456, 861

\bibitem[\protect\citeauthoryear{{Richard} et~al.}{{Richard}
  et~al.}{2008}]{Richard2008a}
{Richard} J., {Stark} D.~P., {Ellis} R.~S., {George} M.~R., {Egami} E., {Kneib}
  J.,  {Smith} G.~P., 2008, \apj, 685, 705

\bibitem[\protect\citeauthoryear{{Richards} et~al.}{{Richards}
  et~al.}{2002}]{Richards2002a}
{Richards} G.~T., {Vanden Berk} D.~E., {Reichard} T.~A., {Hall} P.~B.,
  {Schneider} D.~P., {SubbaRao} M., {Thakar} A.~R.,  {York} D.~G., 2002, \aj,
  124, 1

\bibitem[\protect\citeauthoryear{{Rollinde} et~al.}{{Rollinde}
  et~al.}{2005}]{Rollinde2005a}
{Rollinde} E., {Srianand} R., {Theuns} T., {Petitjean} P.,  {Chand} H., 2005,
  \mnras, 361, 1015

\bibitem[\protect\citeauthoryear{{Schirber}, {Miralda-Escud{\'e}}, \&
  {McDonald}}{{Schirber} et~al.}{2004}]{Schirber2004a}
{Schirber} M., {Miralda-Escud{\'e}} J.,  {McDonald} P., 2004, \apj, 610, 105

\bibitem[\protect\citeauthoryear{{Scott} et~al.}{{Scott}
  et~al.}{2000}]{Scott2000a}
{Scott} J., {Bechtold} J., {Dobrzycki} A.,  {Kulkarni} V.~P., 2000, \apjs, 130,
  67

\bibitem[\protect\citeauthoryear{{Scott} et~al.}{{Scott}
  et~al.}{2004}]{Scott2004a}
{Scott} J.~E., {Kriss} G.~A., {Brotherton} M., {Green} R.~F., {Hutchings} J.,
  {Shull} J.~M.,  {Zheng} W., 2004, \apj, 615, 135

\bibitem[\protect\citeauthoryear{{Shapley} et~al.}{{Shapley}
  et~al.}{2006}]{Shapley2006a}
{Shapley} A.~E., {Steidel} C.~C., {Pettini} M., {Adelberger} K.~L.,  {Erb}
  D.~K., 2006, \apj, 651, 688

\bibitem[\protect\citeauthoryear{{Sijacki}, {Springel}, \&
  {Haehnelt}}{{Sijacki} et~al.}{2009}]{Sijacki2009a}
{Sijacki} D., {Springel} V.,  {Haehnelt} M.~G., 2009, \mnras, 400, 100

\bibitem[\protect\citeauthoryear{{Songaila}}{{Songaila}}{2004}]{Songaila2004a}
{Songaila} A., 2004, \aj, 127, 2598

\bibitem[\protect\citeauthoryear{{Songaila} \& {Cowie}}{{Songaila} \&
  {Cowie}}{2010}]{Songaila2010a}
{Songaila} A.,  {Cowie} L.~L., 2010, \apj, 721, 1448

\bibitem[\protect\citeauthoryear{{Songaila} et~al.}{{Songaila}
  et~al.}{1999}]{Songaila1999a}
{Songaila} A., {Hu} E.~M., {Cowie} L.~L.,  {McMahon} R.~G., 1999, \apjl, 525,
  L5

\bibitem[\protect\citeauthoryear{{Springel}}{{Springel}}{2005}]{Springel2005a}
{Springel} V., 2005, \mnras, 364, 1105

\bibitem[\protect\citeauthoryear{{Springel} et~al.}{{Springel}
  et~al.}{2005}]{Springel2005b}
{Springel} V. et~al., 2005, \nat, 435, 629

\bibitem[\protect\citeauthoryear{{Srbinovsky} \& {Wyithe}}{{Srbinovsky} \&
  {Wyithe}}{2010}]{Srbinovsky2010a}
{Srbinovsky} J.~A.,  {Wyithe} J.~S.~B., 2010, Publications of the Astronomical
  Society of Australia, 27, 110

\bibitem[\protect\citeauthoryear{{Stark} et~al.}{{Stark}
  et~al.}{2007}]{Stark2007b}
{Stark} D.~P., {Ellis} R.~S., {Richard} J., {Kneib} J., {Smith} G.~P.,
  {Santos} M.~R., 2007, \apj, 663, 10

\bibitem[\protect\citeauthoryear{{Steidel} et~al.}{{Steidel}
  et~al.}{1999}]{Steidel1999a}
{Steidel} C.~C., {Adelberger} K.~L., {Giavalisco} M., {Dickinson} M.,
  {Pettini} M., 1999, \apj, 519, 1

\bibitem[\protect\citeauthoryear{{Storrie-Lombardi} et~al.}{{Storrie-Lombardi}
  et~al.}{1994}]{Storrie1994a}
{Storrie-Lombardi} L.~J., {McMahon} R.~G., {Irwin} M.~J.,  {Hazard} C., 1994,
  \apjl, 427, L13

\bibitem[\protect\citeauthoryear{{Telfer} et~al.}{{Telfer}
  et~al.}{2002}]{Telfer2002a}
{Telfer} R.~C., {Zheng} W., {Kriss} G.~A.,  {Davidsen} A.~F., 2002, \apj, 565,
  773

\bibitem[\protect\citeauthoryear{{Theuns} et~al.}{{Theuns}
  et~al.}{1998}]{Theuns1998a}
{Theuns} T., {Leonard} A., {Efstathiou} G., {Pearce} F.~R.,  {Thomas} P.~A.,
  1998, \mnras, 301, 478

\bibitem[\protect\citeauthoryear{{Tytler}}{{Tytler}}{1987}]{Tytler1987a}
{Tytler} D., 1987, \apj, 321, 69

\bibitem[\protect\citeauthoryear{{Tytler} et~al.}{{Tytler}
  et~al.}{2004}]{Tytler2004a}
{Tytler} D. et~al., 2004, \apj, 617, 1

\bibitem[\protect\citeauthoryear{{Utsumi} et~al.}{{Utsumi}
  et~al.}{2010}]{Utsumi2010a}
{Utsumi} Y., {Goto} T., {Kashikawa} N., {Miyazaki} S., {Komiyama} Y.,
  {Furusawa} H.,  {Overzier} R., 2010, \apj, 721, 1680

\bibitem[\protect\citeauthoryear{{Vogt} et~al.}{{Vogt}
  et~al.}{1994}]{Vogt1994a}
{Vogt} S.~S. et~al., 1994, in Society of Photo-Optical Instrumentation
  Engineers (SPIE) Conference Series, Vol. 2198, {D.~L.~Crawford \&
  E.~R.~Craine} , ed, Society of Photo-Optical Instrumentation Engineers (SPIE)
  Conference Series, p. 362

\bibitem[\protect\citeauthoryear{{Williger} et~al.}{{Williger}
  et~al.}{1994}]{Williger1994a}
{Williger} G.~M., {Baldwin} J.~A., {Carswell} R.~F., {Cooke} A.~J., {Hazard}
  C., {Irwin} M.~J., {McMahon} R.~G.,  {Storrie-Lombardi} L.~J., 1994, \apj,
  428, 574

\bibitem[\protect\citeauthoryear{{Wyithe} \& {Loeb}}{{Wyithe} \&
  {Loeb}}{2004}]{Wyithe2004a}
{Wyithe} J.~S.~B.,  {Loeb} A., 2004, \nat, 432, 194

\bibitem[\protect\citeauthoryear{{Wyithe} \& {Bolton}}{{Wyithe} \&
  {Bolton}}{2010}]{Wyithe2010a}
{Wyithe} S.,  {Bolton} J.~S., 2010, ArXiv e-prints

\bibitem[\protect\citeauthoryear{{Yoshida} et~al.}{{Yoshida}
  et~al.}{2006}]{Yoshida2006a}
{Yoshida} M. et~al., 2006, \apj, 653, 988

\end{thebibliography}

\newpage
\clearpage
\newpage

\appendix

\section[]{Detailed discussion of the error analysis}

In this Appendix we show how varying some of
the parameters in Table~\ref{tab:err_budget}
affects the scatter in $\log(\Gamma_{\rm bkg})$
and induce a systematic bias. By modelling the
effects of all these parameters, an expected
sightline-to-sightline scatter, $\sigma _{exp}$,
and bias, $\epsilon _{exp}$, can be calculated
for each individual quasar. The former
contributes to our quoted errors in $\log(\Gamma_{\rm bkg})$,
and the latter is used such that
$\log(\Gamma^{\rm corr}_{\rm bkg}) = \log(\Gamma^{\rm raw}_{\rm bkg}) - \epsilon _{exp}$.
All values of $\log(\Gamma_{\rm bkg})$ quoted in
Table~\ref{tab:results} have been corrected for
these biases. For most of the spectra, the
combined correction for all the sources of error
is $\epsilon _{exp} \lesssim 0.2$ dex.

\begin{figure}
	\centering
   \includegraphics[width=8.0cm]{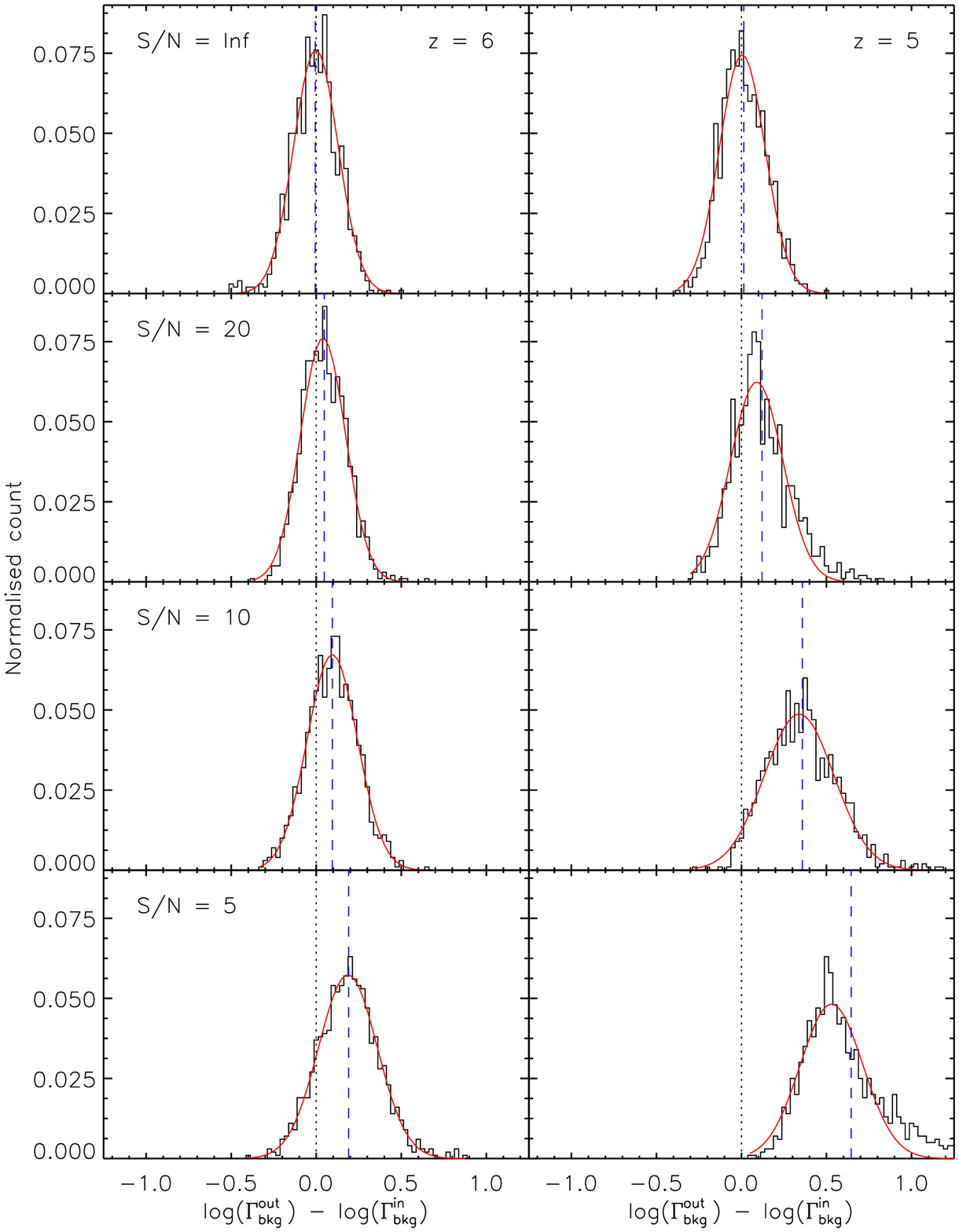}
	\caption{
	Distribution of the estimated values of
	$\log(\Gamma_{\rm bkg})$ as a function of $S/N$
	for 1000 simulated spectra with HIRES resolution.
	A	proximity region size of 10~Mpc is assumed.
	Apart from the luminosity of the quasar and the
	peculiar velocities of the gas, there	are no
	other sources of error. The dotted line marks
	the input value, while the blue dashed line marks
	the mean of the simulated data set. The estimated
	values of $\log(\Gamma_{\rm bkg})$ are well fit
	by a Gaussian.
	}
	\label{fig:hires_fzn}
\end{figure}

\begin{figure}
	\centering
   \includegraphics[width=8.0cm]{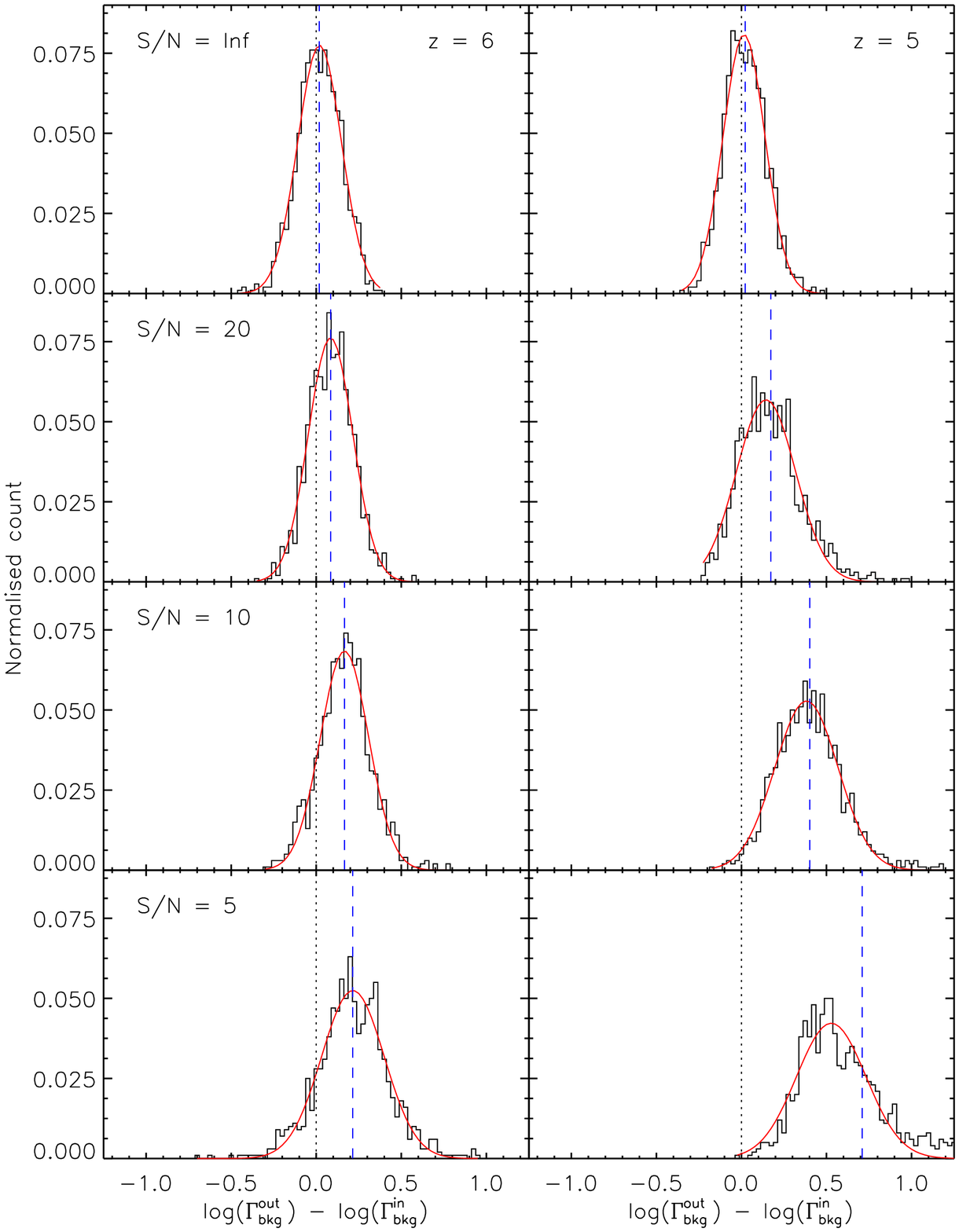}
	\caption{
	Same as Fig.~\ref{fig:hires_fzn} but for simulated
	data with MIKE velocity resolution.
	}
	\label{fig:mike_fzn}
\end{figure}

\begin{figure}
	\centering
   \includegraphics[width=8.0cm]{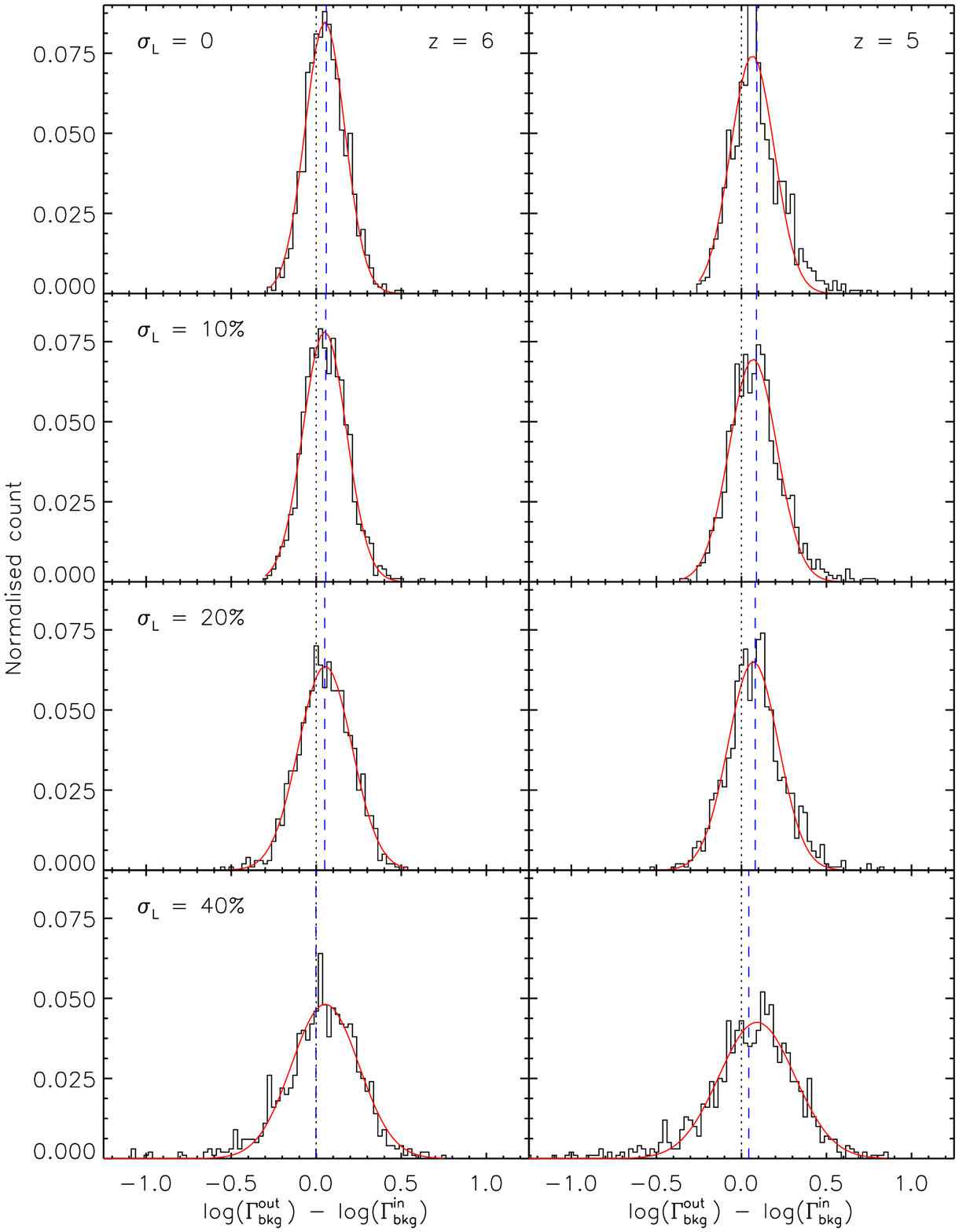}
	\caption{
	Expected distribution of errors in $\log(\Gamma_{\rm bkg})$
	for various errors in the luminosity, assuming a
	fixed	$S/N$ of 20, and HIRES resolution.
	}
	\label{fig:hires_lumerr}
\end{figure}

\begin{figure}
	\centering
   \includegraphics[width=8.0cm]{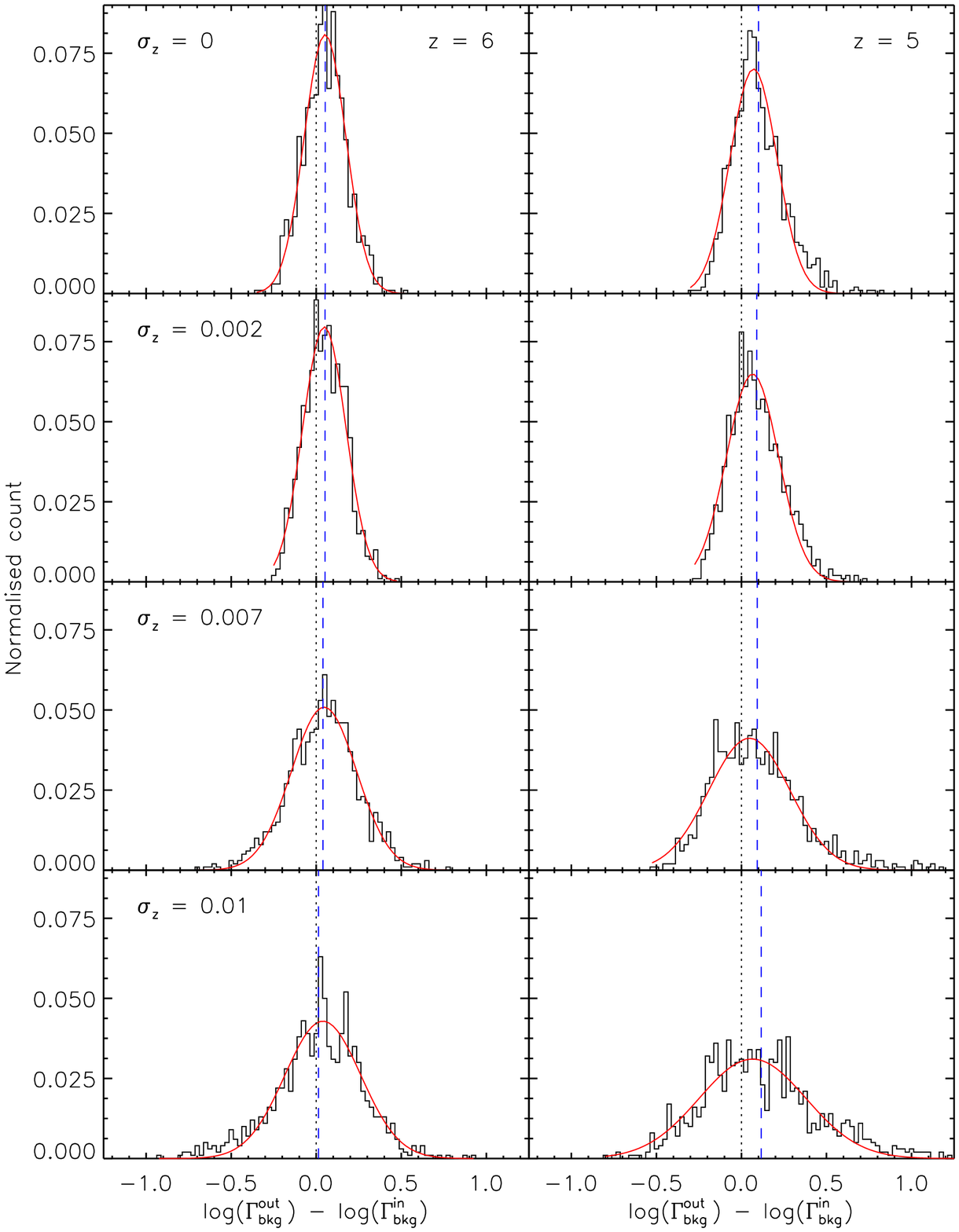}
	\caption{
	Expected distribution of errors in $\log(\Gamma_{\rm bkg})$
	for various errors in the redshift, for an assumed
	$S/N$ of 20, and HIRES resolution. The top panels
	assume no redshift error. The next two panels down
	represent CO and \MgII redshifts, respectively,
	and the	bottom panels represent the errors on
	redshifts determined from the onset of the \Lya
	forest.
	}
	\label{fig:hires_zerr}
\end{figure}

\begin{figure}
	\centering
   \includegraphics[width=8.0cm]{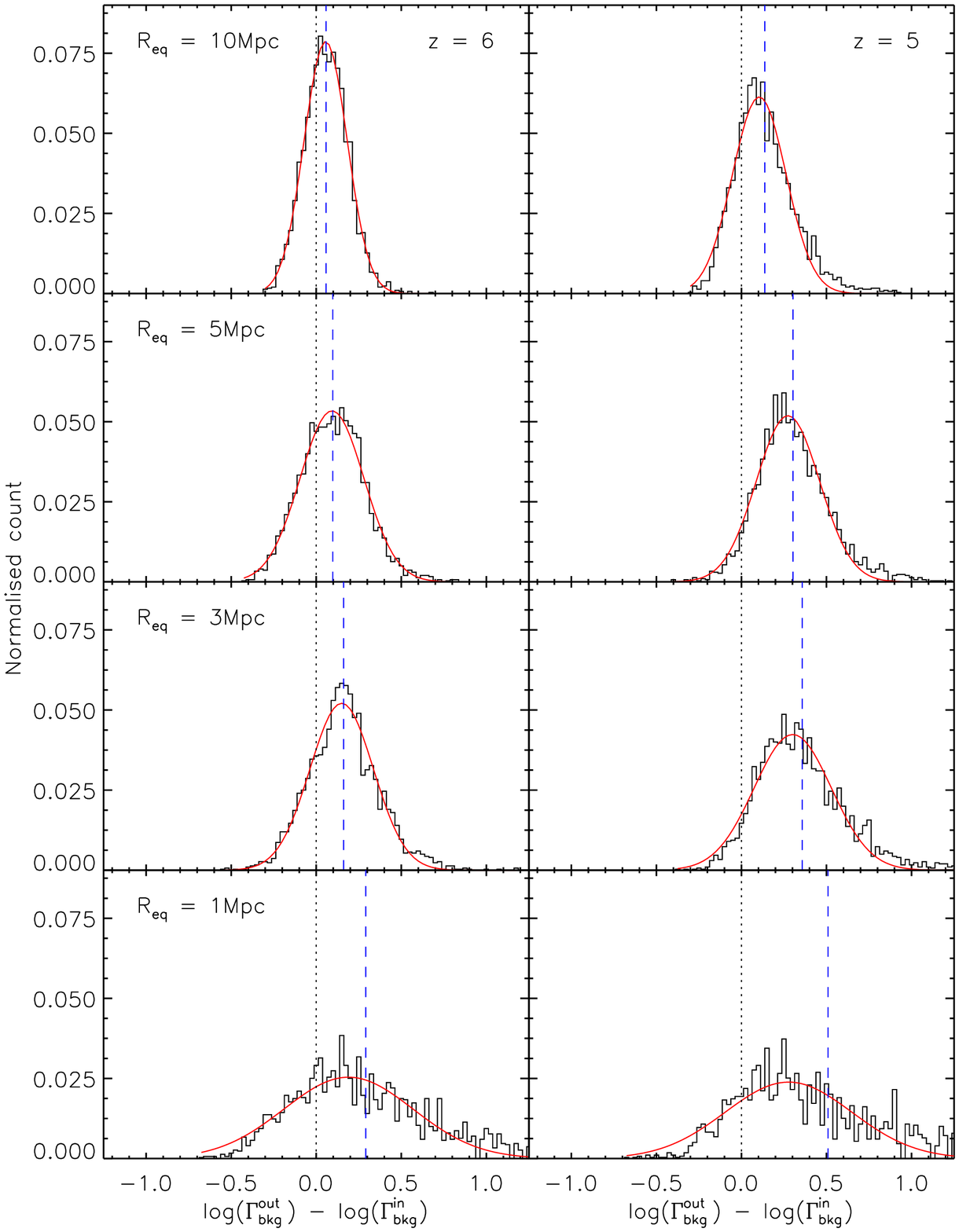}
	\caption{
	Expected distribution of errors in $\log(\Gamma_{\rm bkg})$
	for various input proximity region sizes
	(i.e. luminosity of the quasar), for a fixed
	$S/N$ of 20, and HIRES resolution.
	}
	\label{fig:hires_req}
\end{figure}

\begin{figure}
	\centering
   \includegraphics[width=8.0cm]{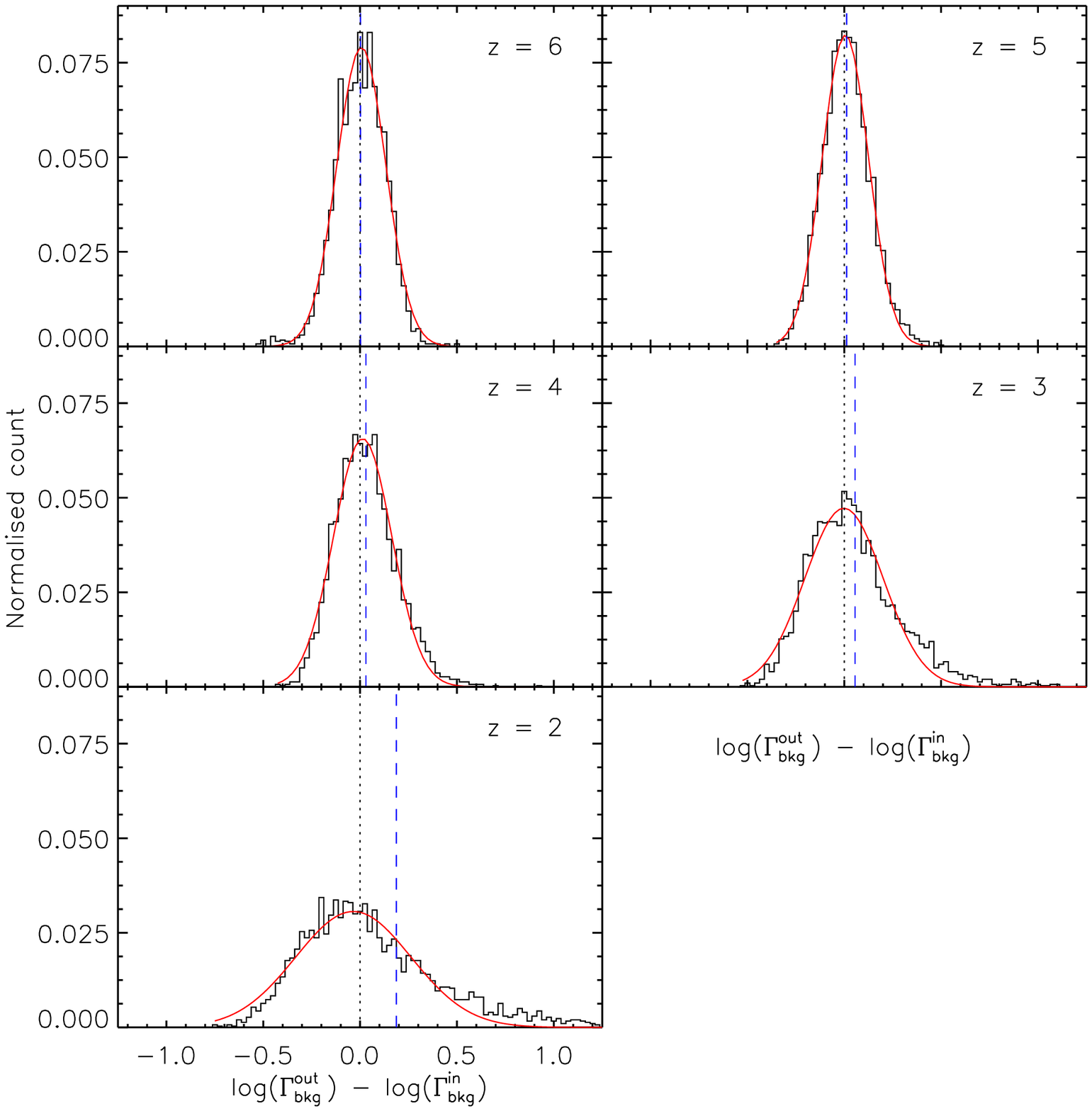}
	\caption{
	Expected distribution of errors in $\log(\Gamma_{\rm bkg})$
	for various quasar redshifts, for noiseless	spectra
	(i.e. $S/N = \infty$), and at HIRES resolution.
	Even for noiseless data, the method struggles
	at lower redshift. The distribution loses its
	Gaussian symmetry, although the mode of the
	distribution does maintain its position close to
	the input value (see also \citet{DallAglio2008a}).
	}
	\label{fig:hires_zpdf}
\end{figure}

\end{document}